\documentclass[a4paper,11pt]{article}

\usepackage{latexsym}
\usepackage[dvips]{graphics}
\usepackage{cite}

\newtheorem{theorem}{Theorem}
\newtheorem{lemma}[theorem]{Lemma}
\newtheorem{coro}[theorem]{Corollary}
\newtheorem{prop}[theorem]{Proposition}

\begin{document}

\def\abstract#1{\begin{center}{\bf ABSTRACT}\end{center}
\par #1}
\def\title#1{\begin{center}{\large {#1}}\end{center}}
\def\author#1{\begin{center}\textsc{#1}\end{center}}
\def\address#1{\begin{center}\textit{#1}\end{center}}

\def\pubnum{00--13}

\begin{titlepage}
\hfill
\parbox{6cm}{{YITP--\pubnum} \par March 2000}
\parbox{6cm}{}
\par
\vspace{1.5cm}
\begin{center}
\Large
Locally U(1)$\times$U(1) Symmetric Cosmological Models: \\ 
Topology and Dynamics
\end{center}
\vskip 1cm
\author{Masayuki TANIMOTO\footnote{JSPS Research Fellow. 
Electronic mail: tanimoto@yukawa.kyoto-u.ac.jp}}

\address{Yukawa Institute for Theoretical Physics,
        Kyoto University, Kyoto 606-8502, Japan.}
\vskip 1 cm

\abstract{We show examples which reveal influences of spatial topologies
  to dynamics, using a class of spatially {\it closed} inhomogeneous
  cosmological models. The models, called the {\it locally
    U(1)$\times$U(1) symmetric models} (or the {\it generalized Gowdy
    models}), are characterized by the existence of two commuting
  spatial {\it local} Killing vectors. For systematic investigations we
  first present a classification of possible spatial topologies in this
  class. We stress the significance of the locally homogeneous limits
  (i.e., the Bianchi types or the `geometric structures') of the models.
  In particular, we show a method of reduction to the natural reduced
  manifold, and analyze the equivalences at the reduced level of the
  models as dynamical models. Based on these fundamentals, we examine
  the influence of spatial topologies on dynamics by obtaining
  translation and reflection operators which commute with the dynamical
  flow in the phase space. }

\end{titlepage} \addtocounter{page}{1}

\def\R{{\bf R}}
\def\Z{{\bf Z}}
\def\x{{\bf x}}
\def\del{\partial}
\def\Lap{\bigtriangleup}
\def\Tr{{\rm Tr}}
\def\^{\wedge}
\def\goinf{\rightarrow\infty}
\def\goes{\rightarrow}
\def\bm{\boldmath}
\def\-{{-1}}
\def\inv{^{-1}}
\def\sqr{^{1/2}}
\def\isqr{^{-1/2}}

\def\reff#1{(\ref{#1})}
\def\vb#1{{\partial \over \partial #1}} 
\def\vbrow#1{{\partial/\partial #1}} 
\def\Del#1#2{{\partial #1 \over \partial #2}}
\def\Dell#1#2{{\partial^2 #1 \over \partial {#2}^2}}
\def\Dif#1#2{{d #1 \over d #2}}
\def\Lie#1{ {\cal L}_{#1} }
\def\diag#1{{\rm diag}(#1)}
\def\abs#1{\left | #1 \right |}
\def\rcp#1{{1\over #1}}
\def\paren#1{\left( #1 \right)}
\def\brace#1{\left\{ #1 \right\}}
\def\bra#1{\left[ #1 \right]}
\def\angl#1{\left\langle #1 \right\rangle}
\def\vector#1#2#3{\paren{\begin{array}{c} #1 \\ #2 \\ #3 \end{array}}}
\def\svector#1#2{\paren{\begin{array}{c} #1 \\ #2 \end{array}}}
\def\matrix#1#2#3#4#5#6#7#8#9{
        \left( \begin{array}{ccc}
                        #1 & #2 & #3 \\ #4 & #5 & #6 \\ #7 & #8 & #9
        \end{array}     \right) }
\def\smatrix#1#2#3#4{ \left( \begin{array}{cc} #1 & #2 \\ #3 & #4
        \end{array}  \right) }

\def\GL#1{{\rm GL}(#1)}
\def\SL#1{{\rm SL}(#1)}
\def\PSL#1{{\rm PSL}(#1)}
\def\O#1{{\rm O}(#1)}
\def\SO#1{{\rm SO}(#1)}
\def\IO#1{{\rm IO}(#1)}
\def\ISO#1{{\rm ISO}(#1)}
\def\U#1{{\rm U}(#1)}
\def\SU#1{{\rm SU}(#1)}

\def\mcg#1{{\rm mcg}(#1)}
\def\omcg#1{{\rm mcg}_+(#1)}

\def\diffeos{diffeomorphisms}
\def\diffeo{diffeomorphism}
\def\Teich{{Teichm\"{u}ller }}
\def\Poin{{Poincar\'{e}}}

\def\Gam{\mbox{$\Gamma$}}
\def\d{{\rm d}}
\def\VII#1{\mbox{VII${}_{#1}$}}
\def\VI#1{\mbox{VI${}_{#1}$}}
\def\Nil{{\rm Nil}}
\def\Sol{{\rm Sol}}
\def\F{{\cal F}}

\def\hh{{h}}
\def\gg{{\rm g}}
\def\uh#1#2{\hh^{#1#2}}
\def\dh#1#2{\hh_{#1#2}}
\def\ug#1#2{\gg^{#1#2}}
\def\dg#1#2{\gg_{#1#2}}
\def\uug#1#2{\tilde{\gg}^{#1#2}}
\def\udg#1#2{\tilde{\gg}_{#1#2}}
\def\udh#1#2{\tilde{\hh}_{#1#2}}

\def\MM{\four{M}}
\def\uMM{\four{\tilde{M}}}
\def\Mtil{\tilde M}

\def\s#1{\sigma^{#1}}
\def\dug#1#2{g_{#1}{}^{#2}}
\def\Conj{{\rm Conj}}

\def\wa{\!\!\!\!&=&\!\!\!\!}
\def\wb{\!\!\!\!&\equiv &\!\!\!\!}
\def\nd{\noindent}
\def\D{{\cal D}} 

\def\UU{\mbox{$\U1\times \U1$}}
\def\LUU{{${\cal LU}{}^2$}}
\def\LUUs{\LUU-symmetric}
\def\Diff#1{{\rm Diff}(#1)} 
\def\Ziff#1{{\rm Diff}_0(#1)}

\def\endofproofmark{\hspace{1em} ${}_\Box$}
\def\proofmark{{\it Proof}:\hspace{1em}}
\def\defmark{{\bf Definition}\hspace{1em}}
\def\defsmark{{\bf Definitions}\hspace{1em}}
\def\notemark{{\bf Note}\hspace{1em}}
\def\remarkmark{{\bf Remark}\hspace{1em}}
\def\conventionmark{{\bf Convention}\hspace{1em}}

\def\q{{\rm q}}

\def\bP{{\bar P}}
\def\bQ{{\bar Q}}
\def\bR{{\bar R}}
\def\bgamma{{\bar \gamma}}
\def\blambda{{\bar \lambda}}

\def\GI{G_\mathrm{I}}
\def\GIZ{\GI^{\angl2}}
\def\GII{G_\mathrm{II}}
\def\GIIZ{\GII^{\angl2}}
\def\Gsix{G_{\mathrm{VI}_0}}
\def\GsixZ{\Gsix^{\angl2}}
\def\Gseven{G_{\mathrm{VII}_0}}
\def\GsevenZ{\Gseven^{\angl2}}
\def\HI{H_\mathrm{I}}
\def\HIZ{\HI^{\angl2}}
\def\HII{H_\mathrm{II}}
\def\HIIZ{\HII^{\angl2}}
\def\Hsix{H_{\mathrm{VI}_0}}
\def\HsixZ{\Hsix^{\angl2}}
\def\Hseven{H_{\mathrm{VII}_0}}
\def\HsevenZ{\Hseven^{\angl2}}
\def\SLZ{\mathrm{SL}^{\angl2}}
\def\HZ{H^{\angl2}}
\def\shortPQ{{x\goes -x}}

\section{Introduction}
\label{sec:intro}

Inhomogeneous spacetime models generally inherit from the full
relativistic dynamics the strong nonlinearity in straightest ways, so
that a moderately simple inhomogeneous model can be a good testing
ground to obtain an insight toward understandings of generic properties
of the relativistic dynamics. One of the best-known such models is the
{\it Gowdy model}, where, first of all, two commuting spatial Killing
vectors are assumed. By this assumption the spatial manifold is
simplified to a one dimensional reduced manifold. Second, it is assumed
that the spatial manifold is compact without boundary, i.e., {\it
  closed}. This assumption is favorable in view of the fact that no
ambiguities occur in the boundary conditions once a topology is
specified. A closed space is physically natural, as well, due to, e.g.,
the finiteness of the gravitational action.

A striking property resulting from these two assumptions is the fact
that the possible spatial topologies are restricted. When Gowdy models
first appeared in 1971 \cite{Gow1} as global generalizations of
cylindrical plane wave solutions, only two spatial topologies,
$S^2\times S^1$ and $S^3$, were considered. After a few years from the
first paper, Gowdy \cite{Gow2} became aware of the fact that any closed
three dimensional Riemannian manifold which admits two commuting Killing
vectors is homeomorphic to one of $T^3$, $S^2\times S^1$, $S^3$ or a
lens space $L(p,q)$. So, he added $T^3$ model to his
consideration. (Since lens spaces are finitely
covered by the sphere $S^3$, he argued little about lens space models.)
This class of solutions or cosmological models consists of the original
version of the Gowdy models.

Recently, it has been pointed out \cite{T,Ren} that a natural
generalization is possible, where the two commuting Killing vectors are
{\it local}, i.e., the Killing vectors are defined in a neighborhood of
every point but are not necessarily globally defined. (More precise
definition of local Killing vectors will be presented in the next
section.) This generalization gives us a set of new favorable models to
test the dynamical properties of relativity, especially, as we will see,
in connection with the spatial topology.

Part of the motivation of this work comes from a plan to systematically
investigate how the spatial topology influence the dynamics of a
spacetime. In fact, the relativistic dynamics of a spacetime often seems
to be influenced by its spatial topology. For example, we know the
recollapsing conjecture \cite{BGT,LW,B}, where it is claimed that the
well-known recollapsing property of a positive curvature (topologically
$S^3$) homogeneous and isotropic cosmological model continues to hold
also when the spacetime is inhomogeneous (if only appropriate energy and
pressure conditions are fulfilled). Thus we may interpret the
recollapsing property is a direct consequence derived from the spatial
topology.

As another example, note (e.g., \cite{RS}) that the dynamics of a vacuum
homogeneous universe can be well characterized by the ``potential''
(given by the spatial scalar curvature), which can be determined if we
know the Bianchi type. The local dynamics of a compactified (locally)
homogeneous model can also be understood in the same manner (see
\cite{KTH,TKH1,TKH2}). Moreover in most cases, the corresponding Bianchi
type is unique if the spatial manifold is closed. Thus we can say
that the spatial topology determines the dynamical behavior of the
universe at least at the locally homogeneous limit.

How about when the spacetime becomes inhomogeneous?  Can one, as in the
locally homogeneous cases, characterize the dynamics of an inhomogeneous
universe by the spatial topology? When the spacetimes become
inhomogeneous distinctions in local properties will obviously disappear,
as the Einstein field equations can be written in the same form,
irrespective of the spatial topology. However, the boundary conditions
may be different in general. So, we can naturally expect that the
distinctions in the dynamical properties due to the differences in the
spatial topologies manifest themselves in their \textit{global,
  discrete, statistical, or average} properties. (Note that the
recollapsing property mentioned above itself is also regarded as a
global property.)  It seems fascinating to find such properties. Also,
such an investigation might provide us with some clues to the topology
of our universe.

In this paper we show, after some preparations availing ourselves of the
generalized Gowdy models, an example of such properties using those
models. More precise contents and contributions should be the following.

We first reveal all the possible spatial topologies of the generalized
Gowdy models and classify them. In particular, we categorize all the
models into two kinds, then subdivide each of them into finite types
each of which has a correspondence to a particular type of locally
homogeneous manifolds. Although we show the best classification we have,
which will require further considerations (in fact, the number of the
possible topologies is infinite), the set of the (finite) types
mentioned above will be found to present the most relevant set of
classes in an appropriate sense. (To be precise, however, some of the
classes can have a discrete parameter.) While by our systematic analysis
we find no new topologies other than those in the literature
\cite{Gow1,Gow2,T,Ren}, one of our contributions is that we first
asserts that those models do exhaust all the possible ones.

Then we will concentrate on the models belonging to one kind, called the
first kind, where each spatial manifold is a $T^2$-bundle over
$S^1$. For these models we argue how we can represent the metric, and
discuss the question of {\it natural reduction} (to $S^1$ as the spatial
part). As a result of the reduction, we find that some models give rise
to the same reduced Einstein equations with the same boundary conditions
for their metric functions. We say that such two models are {\it
  dynamically equivalent} to each other. The dynamical equivalence
greatly diminishes the number of representative models (or topologies)
we should consider. The basic idea for the reductions is the same as the
one presented in Ref.\cite{T}, while the dynamical equivalence is first
introduced in this paper. All of these settings are an indispensable
step for the subsequent study of the models.

Finally, motivated by the reason explained above, we consider varieties
of spatial \textit{translation} and \textit{reflection} symmetries, and
then ask if these symmetries imposed on initial data sets are preserved
in time or not for each reduced model. In fact, we find remarkable
differences in the reflections.

These are done in Secs.\ref{sec:TG} to \ref{sec:sym}.  Section
\ref{sec:summary} is devoted to a summary and comments. In particular,
comments on AVTD behavior and the local $\U1$ models are made. Appendix
A gives a summary of the standard description of the Gowdy models with
generalizations. Appendix B gives a summary of a calculation for the
compactification of Sol geometry.

We adopt the abstract index notation \cite{Wa}, that is, we use small
Latin indices $a,b,\cdots$ {\it not} to denote components but to
represent the type of a tensor explicitly. To denote components we use
Greek indices $\mu,\nu\cdots$ or capital Latin indices
$A,B,\cdots$. Conventionally we write tilde to denote a metric on a
universal cover like $\tilde \dg ab$, while a metric on a quotient space
is simply written as $\dg ab$. In the component representation, however,
these metrics are possibly represented in the same way like $\dg\mu\nu$.
We assume that all the spatial three-manifolds are orientable in this
paper.

\section{Topologies and Geometries of Locally $\U1\times \U1$
 symmetric spaces and spacetimes}
\label{sec:TG}

\begin{figure}[hbtp]
  \begin{center}
    \resizebox{5cm}{!}{\includegraphics{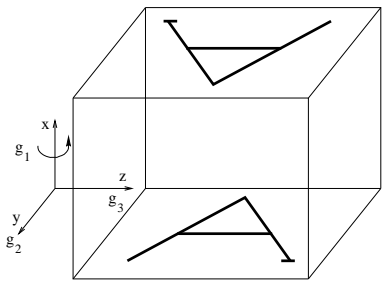}}
    \caption{\small $T^3/\Z_2$, generated from $\R^3$ endowed with the
      Riemannian metric \reff{eq:lk-1}, by three identification generators
      $g_1$, $g_2$, and $g_3$. $g_2$ and $g_3$ are translations along,
      respectively, $y$ and $z$-axes. $g_1$ is the composition of a
      translation along $x$-axis and the rotation in the $y$-$z$
      planes by $\pi$.}
    \label{fig:torus}
  \end{center}
\end{figure}

Let $(M,\dh ab)$ be a Riemannian manifold . Suppose for any point $p$ in
$M$ there exit neighborhood $U$ which admits Killing vector fields, but
these Killing vectors are not necessarily defined on whole $M$. For
example, consider the manifold $\R^3$ with metric
\begin{equation}
  \label{eq:lk-1}
  \d s^2=e^{2\alpha(x)}\d x^2+e^{2\beta(x)}(\d y^2+\d z^2),
\end{equation}
where $\alpha(x)=\alpha(x+2\pi)$ and $\beta(x)=\beta(x+2\pi)$ are real
periodic functions. This Riemannian manifold, denoted as $\tilde M_1$
hereafter, possesses three independent Killing vectors
\begin{equation}
  \label{eq:lk-2}
  \xi_2=\vb y,\; \xi_3=\vb z,\; \xi_4=-z\vb y+y\vb z.
\end{equation}
Introducing identifications in $\tilde M_1$ by infinite actions
generated by 
\begin{eqnarray}
  \label{eq:lk-2.1}
g_1 \equiv e^{\pi \xi_4}e^{2\pi\xi_1} &:& (x,y,z)\goes (x+2\pi,-y,-z),
\nonumber \\
g_2 \equiv e^{2\pi\xi_2} &:& (x,y,z)\goes (x,y+2\pi,z), \nonumber \\
g_3 \equiv e^{2\pi\xi_3} &:& (x,y,z)\goes (x,y,z+2\pi),
\end{eqnarray}
where $\xi_1\equiv \vb x$ is a vector field defined for convenience, we
obtain a closed manifold $M_1$ homeomorphic to $T^3/\Z_2$. (Here, an
exponential of a vector represents the \diffeo\ generated by the
vector.) See Fig.\ref{fig:torus}.  Note that $\xi_2$ (and $\xi_3$) on
the bottom points the opposite direction to that on the top, which fact
tells us that $\xi_2$ and $\xi_3$ cannot be defined on the whole
$M_1$. $\xi_4$ is also incompatible with the identifications by the
translations by $g_2$ and $g_3$. Thus, $M_1$ admits no Killing vectors,
though we can define Killing vectors on a neighborhood of every point.

If we consider a spacetime manifold $(M_1\times\R,\dg ab)$ whose spatial
metric coincides with Eq.\reff{eq:lk-1}, this gives a simplest (but
rather uninteresting) example of a generalized Gowdy model.

\medskip

\nd
\defmark Suppose a Riemannian manifold $(M,\dh ab)$
possesses the following two properties:
\begin{description}
\item[(i)] There exists an open cover $\brace{O_i}$ of $M$ such that
  every $(O_i,\dh ab^{(i)})$ admits $m$ independent Killing vectors
  $\xi_1^{(i)}\cdots\xi_m^{(i)}$, where $\dh ab^{(i)}$ is the natural
  metric on $O_i$ inherited from $\dh ab$.
\item[(ii)]
On each $O_i\cap O_j(\neq\emptyset)$, $\xi^{(i)}$ and $\xi^{(j)}$ are
related through a linear
transformation
\begin{equation}
  \label{eq:lk-3}
  \xi_\alpha^{(i)}=\sum_{\beta=1}^m
  f_\alpha^{(ij)\beta}\xi_\beta^{(j)},\quad
  \alpha=1\sim n,
\end{equation}
where $f_\alpha^{(ij)\beta}$ is a nondegenerate constant matrix (for
fixed $i$ and $j$).
\end{description}
By {\it local Killing vectors on $(M,\dh ab)$}, we mean the collection
of the pairs $(\{O_i\},\{\xi_1^{(i)}\cdots \xi_m^{(i)}\})$ of the cover
and the set of Killing vectors on each patch. For simplicity, we also
simply denote them as $\xi_1\cdots \xi_m$ if we do not have to say on
which patch these Killing vectors are defined.

\medskip From the property (ii), the Killing {\it orbits} are well
defined globally, even though the Killing vectors are local.

\medskip

\nd\defmark A three-dimensional closed Riemannian manifold $(M,\dh ab)$
is called a {\it locally \UU\ symmetric space} (or an {\it \LUUs\ space}
in short) if $(M,\dh ab)$ admits two commuting local Killing vectors
$\bigcup_i (\{O_i\},\{\xi_1^{(i)},\xi_2^{(i)}\})$, i.e.,
$[\xi_1^{(i)},\xi_2^{(i)}]=0$ on every $O_i$. We also impose the
condition that this manifold must have no other local Killing vectors
than those mentioned. We call this condition the {\it genericity
  condition}.

\medskip

As we will see later, the Killing orbits for this kind of space are flat
2-tori if the local Killing vectors are nondegenerate. We can choose
each patch $O_i$ in the above definition as a regular neighborhood of
such a $T^2$ (i.e., a neighborhood such that the boundaries coincides
with another orbits), since the two commuting Killing vectors can be
defined on such a neighborhood. In this case, the group \UU\ acts on
this patch as isometries, hence the word ``locally \UU\ symmetric'' for
the whole manifold $(M,\dh ab)$.  However, this terminology is rather
lengthy to use frequently, so we also use ``\LUUs'' or more simply
``\LUU-'' in this paper.

 \medskip

 \nd\defmark A spacetime manifold $(M\times\R,\dg ab)$ is called a {\it
   locally \UU\ symmetric spacetime} (or a {\it generalized Gowdy
   spacetime}) if the spatial manifold $M$ is closed and the spacetime
 admits two commuting spatial local Killing vectors. The genericity
 condition is understood, as in the case of the \LUUs\ spaces.

\medskip

Our first task is to determine all the possible topologies for the
locally \UU\ symmetric spaces. It is convenient to consider separately
the case where the local Killing vectors degenerate on some points and
the case where they do not on any points. We call the latter
(nondegenerate) type of manifolds the {\it first kind}, and the former
(degenerate) ones the {\it second kind}.

\subsection{The first kind}
\label{sub:1-1}

\begin{lemma}
  \label{le:T2}
  If $(M,\dh ab)$ is an orientable locally \UU\ symmetric space of the
  first kind, then all Killing orbits must be closed and homeomorphic
  to $T^2$.
\end{lemma}

\def\de#1#2{\eta_{#1#2}} 

\proofmark Let $(O,\de ab)$ be a Killing orbit, where $\de ab$ is the
induced metric on the orbit from $\dh ab$.  Since the Killing orbit
possesses two commuting Killing vectors, $\de ab$ is a flat
metric. Since $M$ is orientable, $O$ is homeomorphic to one of $\R^2$,
$\R\times S^1$, or $T^2$. First, let us assume $O\simeq \R^2$. Consider
a geodesic $l$ in $(O,\de ab)$, and consider a point sequence $\{q_i\}$
such that all $q_i$ are on $l$ and $q_i$ and $q_{i+1}$ are unit distant
with respect to $\de ab$. In $M$ the sequence $\{q_i\}$ must have a
converging subsequence $\{p_i\}$, since $M$ is compact. Let $V_\epsilon$
be the neighborhood of the limit point $p_\infty$ with radius
$\epsilon$, i.e., $|p-p_\infty|<\epsilon$ for all $p\in V_\epsilon$,
where $|\cdot|$ is the geodesic distance with respect to $\dh
ab$. $V_\epsilon$ contains all $p_i$ $(i>N_\epsilon)$ for an integer
$N_\epsilon$, so that for any $p_i$ and $p_j$, $(i,j>N_\epsilon)$,
$|p_i-p_j|<2\epsilon$ as a result of the triangle inequality. On the
other hand, the geodesic $l$ is a ``straight line'' in the Euclid plane
$(\R^2,\de ab)$, so that the distant between any two points $p_i$ and
$p_j$ $(i\neq j)$ in $O$ is larger than or equal to unity.  Hence we can
choose a sequence of ``isometric disconnected neighborhoods'' $\{U_i\}$
in $M$ such that $p_i\in U_i$, $S_i\cap S_j=\emptyset$, where $S_i\equiv
U_i\cap O$, and there exist an isometry $\phi_{ij}:\;
(p_i,S_i,U_i)\goes(p_j,S_j,U_j)$ for all $i,j$. Since $S_i\cap
S_j=\emptyset$ and $|p_i-p_j|<2\epsilon$, we must have the consequence
$p_j\not\in S_i$ but $p_j\in U_i$ for a sufficiently small $\epsilon$
(i.e., for sufficiently large $i$ and $j$).  Moreover, this implies that
for an {\it arbitrarily small} $\epsilon'< \epsilon$ there exist two
points in $M$, $p_i$ and $p_j$, such that $|p_i-p_j|<2\epsilon'$,
$p_j\not\in S_i$ but $p_j\in U_i$, and there exist the isometry
$\phi_{ij}:\; p_i\goes p_j$. This in turn implies that there exist a
third motion in $(M,\dh ab)$ off the orbit, which fact is against the
genericity condition. (That is, $\R^2$-orbits can be formed only if
$(M,\dh ab)$ is locally homogeneous.) In the case of $O\simeq \R\times
S^1$, we can choose a geodesic $l$ which is open as in the case of
$O\simeq \R^2$. Repeating a similar argument we conclude that this case
is also against the genericity. Thus, the only possibility is the case
$O\simeq T^2$. \endofproofmark

\begin{theorem}
  \label{th:bundle}
  If $(M,\dh ab)$ is an orientable locally \UU\ symmetric space of the
  first kind, then $M$ is homeomorphic to a $T^2$-bundle over $S^1$, for
  which the local Killing vectors generate the $T^2$-fibers.
\end{theorem}

\proofmark Consider a normal vector field that is everywhere
nonvanishing and normal to the Killing orbit, and consider the
integration curve $c$ of this field passing through an arbitrarily given
point $p\in M$. Let $O$ be the Killing orbit to which $p$ belongs, so
$c$ intersects with $O$ at $p$. If $c$ did not intersect again with $O$,
then it would imply that the total volume of $M$ is infinite, which fact
contradicts with the compactness of $M$. Hence $c$ intersects with $O$
again. This implies that if we regard each Killing orbit as a point, then
$M$ reduces to $S^1$. Since all orbits are homeomorphic to $T^2$ as
in Lemma \ref{le:T2}, $M$ is a $T^2$-bundle over $S^1$. \endofproofmark

\medskip

Topologically, any $T^2$-bundle over $S^1$ can be obtained by first
considering the product $T^2\times I$, where $I=[0,1]$ is the unit
interval, then identifying $T^2\times\{0\}$ and $T^2\times\{1\}$ by the
action of a gluing map $\phi:\; T^2\goes T^2$. Since any two gluing maps
which are homotopic to each other results in topologically the same
manifold, we usually think of the gluing map $\phi$ as an element of the
{\it mapping class group} of $T^2$, $\mcg{T^2}$. Here, the mapping class
group of a manifold $M$ is the group of \diffeos\ of $M$ modulo
\diffeos\ which are homotopic to the identity, $
\mcg{M}=\Diff{M}/\Ziff{M} $. The group $\mcg{T^2}$ is well known as the
{\it modular group}, $\GL{2,\Z}$. However, since we are interested only
in orientable manifolds, we consider the orientation-preserving
mapping class group of the torus, $\omcg{T^2}\simeq\SL{2,\Z}$.  Letting
\begin{equation}
  \label{eq:defA}
  A=\smatrix pqrs\in\SL{2,\Z},
\end{equation}
the fundamental group of the corresponding space is
represented by
\begin{equation}
  \label{eq:pi1}
  \pi_1=\angl{g_1,g_2,g_3;\;
    [g_2,g_3]=1,\;
    g_1 g_2 g_1\inv=g_2^p g_3^r,\;
    g_1 g_3 g_1\inv=g_2^q g_3^s },
\end{equation}
where $g_2$ and $g_3$ are generators of the fiber $T^2$ and $g_1$ is
that of the base $S^1$. We therefore have a natural map $\omega_1:
\SL{2,\Z}\goes W_1$, where $W_1$ is the set of orientable $T^2$-bundles
over $S^1$. While $\omega_1$ is not injective, we have the following
useful theorem, of which proof can be found in Ref.\cite{Sco} (Theorem
5.5 therein);
\begin{theorem}[\cite{Sco}]
  \label{th5.5}
  Let $M$ be the total space of a $T^2$-bundle over $S^1$ with gluing
  map $\phi$, and let $A\in\GL{2,\Z}$ represent the automorphism of
  $\pi_1(T^2)$ induced by $\phi$. Then, $M$ admits a Sol-structure if
  $A$ is hyperbolic, an $E^3$-structure if $A$ is periodic, or a
  Nil-structure otherwise. In particular, if $|\Tr A|>2$, then $A$ is
  hyperbolic, so $M$ admits a Sol-structure.
\end{theorem}

This theorem tells us that every $T^2$-bundle over $S^1$ admits a
``geometric structure'' \cite{Th}, in other words, admits a locally
homogeneous metric. This structure is, as in the above theorem, one of
the three types, $E^3$, Nil, or Sol. Conversely, we can find all the
$T^2$-bundles over $S^1$ from the closed quotients of $E^3$, Nil, and
Sol, so we can refer to known classifications of the closed quotients of
the three geometries to classify $T^2$-bundles over $S^1$. This
procedure simultaneously determines the corresponding geometric
structure for each representative. (To avoid confusions we remark that
locally homogeneous metrics are used just for convenience here to
discuss topologies, but we will see that appropriate inhomogeneous
metrics are obtained by ``relaxing'' such a metric.)

{\it The case of $E^3$}: All the orientable closed manifolds modeled on
$E^3$ are classified into six manifolds \cite{Wo}, $T^3$, $T^3/\Z_2$,
$T^3/\Z_3$, $T^3/\Z_4$, $T^3/\Z_6$, and $T^3/\Z_2\times\Z_2$. All of
these {\it except} the last one can be regarded as $T^2$-bundle over
$S^1$. Associated with these five are the representative elements
$E_\lambda\in\SL{2,\Z}$ $(\lambda=1,2,3,4,6,{\rm respectively})$:
\begin{equation}
  \label{eq:A}
  E_1=\smatrix 1001,\; E_2=\smatrix {-1}00{-1},\; 
  E_3=\smatrix 01{-1}{-1},\; E_4=\smatrix 01{-1}0,\;
  E_6=\smatrix 01{-1}1.
\end{equation}

{\it The case of Nil}: All the orientable closed manifolds modeled on
Nil can also be completely classified (See page 4878 of Ref.\cite{KTH}).
Among them, ones which can be regarded as $T^2$-bundle over $S^1$ are
given by the following two families $N_{\pm 1}(n)\in\SL{2,\Z}$:
\begin{equation}
  \label{eq:Anil}
  N_1(n)=\smatrix {1}{n}{0}{1},\; N_{-1}(n)=\smatrix {-1}{n}{0}{-1},
\end{equation}
where $n$ is a nonzero integer. The parameter $n$ in $N_1(n)$ can be
chosen to be positive, since the relabeling $(g_1,g_3)\goes (g_3,g_1)$
reverses the sign of $n$. For $N_{-1}(n)$, the opposite sign of $n$
corresponds to a distinct topology.

{\it The case of Sol}: As in Theorem \ref{th5.5}, if $|\Tr A|>2$, we
have a $T^2$-bundle over $S^1$ modeled on Sol. All those for $|\Tr
A|\leq 2$ are non-orientable \cite{Sco}, so we will not consider them.
Any two ones with distinct values of $\Tr A$ such that $|\Tr A|>2$ are
not homeomorphic, so one parameter family of $\SL{2,\Z}$ \cite {KTH},
\begin{equation}
  \label{eq:Asol}
  S(n)=\smatrix 01{-1}n,
\end{equation}
where $n$ is an integer such that $|n|>2$, gives a one parameter family
of distinct $T^2$-bundles over $S^1$ modeled on Sol. However, this does
{\it not} implies that the ones modeled on Sol can be classified only by
$\Tr A$ like Eq.\reff{eq:Asol}, since there exist topologically distinct
manifolds with the same $\Tr A$ \cite{Kod}. No complete classification
of the closed manifolds modeled on Sol seems to be known. Nevertheless,
as we will see in the next section, the sequence \reff{eq:Asol} gives
nice representatives as reduced relativistic models.

\subsection{The second kind}
\label{sub:1-2}

\begin{theorem}
  \label{th:lens}
  If $(M,\dh ab)$ is a locally \UU\ symmetric space of the second kind,
  then $M$ is homeomorphic to one of $S^3$, $S^2\times S^1$, or a lens
  space $L(p,q)$.
\end{theorem}

\proofmark Since $M$ is closed, if removing an open neighborhood of the
degenerate points from $M$, the resulting manifold $\hat M$ with
boundaries is compact. Moreover, the boundaries can be chosen so as to
coincide with (regular) Killing orbits. Let $\partial \hat M$ be the
boundaries of $\hat M$ so chosen. Note that in the proof of Lemma
\ref{le:T2}, only the compactness of $M$ is assumed to show that the
Killing orbits are $T^2$. Hence all the Killing orbits for $(\hat
M,\hat\dh ab)$ are also $T^2$. (Here, $\hat\dh ab$ is the restriction of
$\dh ab$ to $\hat M$.) In particular, every connected component of
$\partial\hat M$ is $T^2$. Consider a (nonvanishing) normal vector field
with respect to the Killing orbits in $(\hat M,\hat \dh ab)$. This flow
of the normal vector field defines a unique one-to-one correspondence
between boundary components, since the image of a boundary component by
this flow must end up with another boundary component, due to the
compactness of $\hat M$. Since $M$ and therefore $\hat M$ are assumed to
be connected, $\hat M$ must have only two boundary components and is
naturally homeomorphic to the product $T^2\times I$, where $I\equiv
[0,1]$ is the unit interval. Next, consider the neighborhood of a
degenerate Killing orbit removed. An action of \UU\ can degenerate only
to $\U1$, so a connected set of degenerate points forms a circle, and a
neighborhood of such a circle with boundary forms a solid torus. The
boundary, which is homeomorphic to $T^2$, can again be chosen so as to
coincide with a (regular) Killing orbit. Consider two such
neighborhoods, $V_1$ and $V_2$. Then, the original manifold is recovered
by attaching them to $\hat M$ along the boundaries: $M\simeq V_1\cup\hat
M\cup V_2$.  Note, however, that $V_1\cup \hat M$ is another solid torus
with the \UU\ symmetry. Let us rewrite this manifold as $V_1$. The
original manifold $M$ is now obtained simply by identifying the
boundaries of the two solid tori $V_1$ and $V_2$, $M\simeq V_1\cup V_2$.
Finally, it is a well-known fact (e.g. \cite{Hem}) that the sum of two
solid tori is homeomorphic to one of $S^3$, $S^2\times S^1$, or a lens
space $L(p,q)$.  \endofproofmark

\medskip

Cosmological models based on \LUUs\ spatial manifolds of the second
kind can be thought of as a global generalization of cylindrically
symmetric (plane wave) models. In fact, the symmetry axis of a
cylindrically symmetric space is a degenerate orbit for the \UU\ action
of the cylindrical symmetry. If, as usual, taking this axis as
$z$-coordinate axis, and introducing identifications $z\sim z+z_0$ for a
constant $z_0$, we have the symmetry axis that is a circle. A regular
neighborhood of this circle is a solid torus with the \UU\ action. An
\LUU-manifold $M$ can be obtained by identifying (the boundaries of)
such two \UU\ symmetric solid tori.

Now, it should be remarked that the local Killing vectors for an
\LUU-manifold of the second kind are always globally defined. This is
because this kind of space can also be regarded as the product
$T^2\times I$ (with a metric that degenerates on the boundaries). The
original Gowdy models \cite {Gow2} in fact include all the models based
on $S^3$, $S^2\times S^1$, and the lens spaces $L(p,q)$. We therefore do
not discuss much about these models in subsequent sections.

We end this subsection with a summary of the classification
(e.g. \cite{Hem}) of the lens spaces, for completeness. Consider a
gluing map $\phi:T^2\goes T^2$ to identify the boundaries of two \UU\ 
symmetric solid tori, $\del V_1$ and $\del V_2$.  Since if two gluing
maps are homotopic the resulting manifolds are homeomorphic, it is again
natural to think of an identification map as an element of the mapping
class group, $\phi\in\mcg{T^2}\simeq\GL{2,\Z}$. Let $\svector 10$ and
$\svector 01$ be, respectively, meridian and longitudinal loops of $\del
V_1$ or $\del V_2$. We define the action of $A=\smatrix
prqs\in\GL{2,\Z}$ by the left action to these vectors. Then, the closed
manifold $M$ made by gluing the two solid tori with respect to $A$ is
the lens space $L(p,q)$.  (The topology of $M$ is regardless to $r$ and
$s$, i.e., it is determined only by the mapping of the meridian loop of
$\del V_1$.)  The integers $p$ and $q$ are coprime and we can set
$p\geq0$ (since $L(p,q)\simeq L(-p,q)$). $S^3$ and $S^2\times S^1$
correspond, respectively, to $L(1,q)$ and $L(0,1)$. Lens spaces are
completely classified by the fact that $L(p,q)$ and $L(p',q')$ are
homeomorphic if and only if $p=p'$, and $q\equiv \pm q'\;{\rm mod}\; p$
or $qq'\equiv\pm1\;{\rm mod}\; p$.

\section{Relaxation, Reduction and Dynamical Equivalence}
\label{sec:reduction}

In this section we present an detailed account of the reduction
procedure for the \LUUs\ models of the {\it first kind}, with emphasis
on the significance of the corresponding geometric structures. We will
also obtain useful results, the ``dynamical equivalences''.

First, we remark that since the spacetime manifold is
the product $M\times\R$, where the spatial manifold $M$ is a
($T^2$-)bundle over $S^1$ with fibers generated by the local Killing
vectors, the natural reduced manifold obtained by contracting the
Killing orbits should be $S^1\times\R$. However, as we see bellow, when
the bundle is not (covered by) a direct product, this reduction is not
trivial.

We represent an \LUU-symmetric spacetime (of the first kind)
$(M\times\R,\dg ab)$ with the covering map
\begin{equation}
  \label{eq:covmap}
  \pi:\; (\tilde M\times\R,\udg ab)\goes (M\times\R,\dg ab)=(\tilde
  M\times\R,\udg ab)/\Gamma,
\end{equation}
where $(\tilde M\times\R,\udg ab)$ is the universal covering manifold of
$(M\times\R,\dg ab)$, and $\Gamma$ is a covering group, which acts on
$\tilde M$ discretely. Since the spatial manifold $M$ is a $T^2$-bundle
over $S^1$, $\tilde M$ is homeomorphic to $\R^3$. We can therefore use
globally defined coordinates, say $(x^0,x^1,x^2,x^3)\equiv(t,x,y,z)$, to
represent $\udg ab$. A natural form of the metric admitting two
commuting Killing vectors represented with these coordinates is
$\dg\mu\nu(t,x)dx^\mu dx^\nu$, where Greek indices $\mu,\nu,\cdots$ run
over $0$ to $3$. Moreover, fixing the freedom of \diffeos\ in
$\dg\mu\nu(t,x)dx^\mu dx^\nu$ as much as possible, we can represent the
metric $\udg ab$ as \cite{Gow2}
\begin{equation}
  \label{eq:fullg1metric}
  \d s^2= e^{\gamma/2}(-\d t^2+\d x^2)+ 2N_A\d x^A\d t+
  R e_{AB}\d x^A\d x^B,
\end{equation}
where $\gamma$, $R$, $N_A$ and $e_{AB}$ (Capital indices $A,B,\cdots$
run 2 to 3) are functions of $t$ and $x$. We set $\det e_{AB}=1$ so that
$R$ describes the area of the Killing orbit. This metric is called {\it the
generic metric of the canonical form}.
We will also consider the restricted metric
with $N_A=0$ (the ``two-surface orthogonality'' \cite{Gow2})
\begin{equation}
  \label{eq:g1metric}
  \d s^2= e^{\gamma/2}(-\d t^2+\d x^2)+
  R e_{AB}\d x^A\d x^B,
\end{equation}
which is called {\it the two-surface orthogonal metric of the canonical
  form}. For a standard prescription of Einstein's equations for the
last metric, see Appendix \ref{sec:gc}. (Note that the metric functions
are represented with ``bars'' in this Appendix. We will take this
notation from the next subsection to avoid conflicts with
``noncanonical'' metrics which will appear.)

Both metrics admit two dimensional isometry group consisting of
translations along $y$ and $z$ axes. Note that the $y$-$z$ planes
descend to the $T^2$-fibers after appropriate identifications in each
$y$-$z$ plane. These identifications are generated by two independent
vectors, which we can choose without loss of generality as the unit
coordinate generators, $\vbrow y$ and $\vbrow z$. Then we can naturally
think of the actions of these generators as the representation of the
generators, $g_2$ and $g_3$, of $\pi_1$ into the isometry group;
$g_2\simeq e^{\vb y}$, $g_3\simeq e^{\vb z}$. The remaining generator
$g_1$ must induce a translation along $x$-axis {\it plus} the modular
transformation induced by an element $A=n^A{}_B\in\SL{2,\Z}$. Thus,
$g_1:\; (x,y,z)\goes (x+2\pi, n^A{}_B x^B)$ on the $t=$constant space.
Together with $g_3: (x,y,z)\goes (x,y+2\pi,z)$ and $g_3: (x,y,z)\goes
(x,y,z+2\pi)$, all the relations in the fundamental group \reff{eq:pi1}
are fulfilled. The boundary conditions for the metric
\reff{eq:fullg1metric} or \reff{eq:g1metric} is determined by the
requirement that the action of $g_1$ be an isometry of the metric, which
can be easily found;
\begin{eqnarray}
  \label{eq:g1bc}
  & & \lambda(t,x)=  \lambda(t,x+2\pi) , \; R(t,x)= R(t,x+2\pi),
  \nonumber \\
  & & N_A(t,x)=  n^B{}_{A} N_B(t,x+2\pi) , \;
  e_{AB}(t,x)= n^C{}_{A}n^D{}_{B}e_{CD}(t,x+2\pi).
\end{eqnarray}
Note that the components $e_{AB}$ that describe each $T^2$-fiber are
{\it not} periodic functions in general, so that the reduced manifold
(spanned by $t$ and $x$) cannot be naturally regarded as $S^1\times\R$,
if we represent the metric as Eq.\reff{eq:fullg1metric} or
\reff{eq:g1metric}. The reason why $e_{AB}$ do not automatically become
periodic is that what two independent components in $e_{AB}$ themselves
describe are \Teich parameters for the $T^2$-fiber, rather than moduli
parameters. (For the difference between \Teich and moduli parameters,
see, e.g., Ref.\cite{Th}.)

An idea to remedy the situation is to choose the metric functions so
that they become constant when the spatial metric is at a locally
homogeneous limit. That is, conversely, we ``relax'' a locally
homogeneous metric in an appropriate way to obtain a suitable
metric. Note that a constant function is naturally a function on
$S^1$. As far as considering smooth deformations of the metric, metric
functions chosen in such a way must continue to be (smooth) functions on
$S^1$ even when the metric becomes inhomogeneous (with the \LUU\ 
symmetry).

{\bf The scheme}\hspace{1em} One nice way to realize this idea is to
expand the spatial metric in terms of the (left) invariant one-forms $\s
i$ $(i=1\sim3)$ of a Bianchi group $G$. As we have seen in
Sec.\ref{sub:1-1}, any $T^2$-bundle over $S^1$ admits one of $E^3$, Nil,
or Sol-structure. These geometric structures correspond, respectively,
to Bianchi I(VII${}_0$), II, and VI${}_0$.  ($E^3$ has multiple
correspondences.) The appropriate Bianchi group $G$ is determined from
this correspondence.  Let $\xi_i$ $(i=1\sim3)$ be independent generators
of $G$, that is, $\xi_i$ are Killing vectors of a $G$-invariant
metric. Since $\s i$ are by definition invariant under the action of
$G$, the homogeneous metric is written as $\dh ij\s i\s j$ with
(spatially) constant components $\dh ij$. The Killing vectors of this
homogeneous metric include a commuting pair. Suppose
$[\xi_2,\xi_3]=0$. Let $\chi$ be a vector field such that $\chi$ is
independent of $\xi_2$ and $\xi_3$ at every point and is invariant under
the action generated by $\xi_2$ and $\xi_3$, i.e.,
$\Lie{\xi_2}\chi=\Lie{\xi_3}\chi=0$. Here, $\Lie{}$ represents the Lie
derivative. Since $[\xi_2,\xi_3]=0$, we can choose coordinates ($y$ and
$z$) generated by $\xi_2$ and $\xi_3$. Moreover, since
$\Lie{\xi_2}\chi=\Lie{\xi_3}\chi=0$ implies the commutativity
$[\xi_2,\chi]=[\xi_3,\chi]=0$, we can choose a third coordinate ($x$) as
that of generated by $\chi$. Consider an arbitrary function $f(x)$ that
depends only on $x$ chosen in this way. Then, the level set of $f(x)$ is
invariant under the \diffeos\ generated by $\xi_2$ and $\xi_3$, and it
coincides with the set of the orbits generated by $\xi_2$ and $\xi_3$.
Hence we can write the inhomogeneous metric as $\dh ij(x)\s i\s j$,
which is invariant under the \diffeos\ generated by $\xi_2$ and $\xi_3$,
and inhomogeneous in the desired manner. (Recall that all $\s i$ are
invariant under $\xi_2$ and $\xi_3$.) The metric functions $\dh ij(x)$
should be periodic in $x$ because of the reason explained above. Hence,
a spacetime metric with $\dh ij(t,x)\s i\s j$ as the spatial part
naturally gives rise to a reduction onto $S^1\times\R$. We call this
scheme the {\it relaxation method}. If $\xi_1=\chi$ as in the Bianchi I
case, then this scheme descends to the usual coordinate representation
like Eq.\reff{eq:fullg1metric} or \reff{eq:g1metric}, but otherwise it
is nontrivial. \endofproofmark

Recall that a Bianchi group $G$ is a three-dimensional simply transitive
group acting on a three-dimensional simply connected manifold $\tilde
M$, and therefore we can identify $G$ with $\tilde M$
(e.g. \cite{Wa}). We adhere to this viewpoint in this paper and use the
components of $G$ to represent the coordinates of $\tilde M$, as well.

One comment should be made here. As pointed out in Ref.\cite{T}, all
Bianchi homogeneous spaces except for VIII and IX possess a commuting
pair of Killing vectors. Even for VIII and IX, if considering higher
symmetry there can exist commuting Killing vectors. Moreover, these
homogeneous spaces can be compactified to closed spaces, except for IV
and VI${}_a$ types. Therefore all the locally homogeneous manifolds of
the Bianchi types except IV and VI${}_a$ seems to be locally homogeneous
limits of \LUU-manifolds. However, as we mentioned above, only Bianchi
I, VII${}_0$, II, and VI${}_0$ types can be such a limit for the first
kind manifolds. The second kind manifolds correspond to Bianchi IX (in
case of $M\simeq S^3$ and $L(p,q)$) and the Nariai-Kantowski-Sachs type
(in case of $M\simeq S^2\times S^1$). Thus, {\it Bianchi III, V,
  VII${}_a$, VIII types are missing in the list of possible manifolds as
  the locally homogeneous limits of our \LUU-manifolds}. The reason is
because each orbit of the two commuting local Killing vectors for a
locally homogeneous manifold of these types always do not close. For
example, in the case of Bianchi V, it forms $\R^2$. As we have seen in
the proof of Lemma \ref{le:T2}, such a case is possible only when the
manifold is locally homogeneous. That is, we can say that, as a
corollary to Theorems \ref{th:bundle} and \ref{th:lens}:
\begin{coro}
  Every closed locally homogeneous manifold of Bianchi III, V, VII${}_a$
  and VIII types cannot be relaxed to be inhomogeneous with  \LUU\ 
  symmetry.
\end{coro}
In this sense, these manifolds are ``stiff.''  It is worth noting that a
common feature of them is that they all contain a hyperbolic structure,
$H^2$ or $H^3$.

\bigskip

In the following subsections, we perform the reduction procedure for
each type of Nil, Sol, and $E^3$. The spacetime manifold $M\times\R$
reduces to $S^1\times\R$ for any spatial manifold $M$. One important
feature that is found as a result of the reductions is that, even for
models with topologically distinct spatial manifolds, distinctions at
the reduced level can completely {\it degenerate}. In other words, we
obtain for some models the same set of PDEs (as the reduced Einstein
equations) with the same boundary conditions. For such models the
universal covering spacetime metrics can be represented in exactly the
same form with only the difference contained in the covering group
$\Gamma$. We call that such models are {\it dynamically equivalent} to
each other.

\subsection{Case of Nil}
\label{sec:nil}

The Bianchi II group $G_{{\rm II}}$ is the three-dimensional simply
transitive group with the multiplication rule
\begin{equation}
  \label{eq:GIImulti}
  \vector abc\vector xyz=\vector {a+x}{b+y}{c+z+ay}.
\end{equation}
This group is generated by
\begin{equation}
  \label{eq:II-K}
  \xi_1=\vb x+y\vb z,\; \xi_2=\vb y,\; \xi_3=\vb z.
\end{equation}
The one-forms which are invariant under the action of $G_{{\rm II}}$ are
\begin{equation}
  \label{eq:1-nil}
  \s1=\d x,\; \s2=\d y,\; \s3=\d z-x\d y.
\end{equation}
Since $\xi_2$ and $\xi_3$ commute and they generate coordinates $y$
and $z$, a desired representation of the spatial metric is given by $\dh
ij(x)\s i\s j$.

Before proceeding further, however, we have to clarify a subtle point
about the fact that we can choose another commuting pairs of group
generators. For example, $\xi_1$ and $\xi_3$ also commute to each other,
and the orbits they generate are distinct from those generated by
$\xi_2$ and $\xi_3$. In general, since the commutation rule for linear
combinations of the group generators is given by
\begin{equation}
  \label{eq:nil-tmp1}
  [\alpha^i\xi_i,\beta^j\xi_j]=
  \alpha^i\beta^j[\xi_i,\xi_j]=-(\alpha^1\beta^2-\alpha^2\beta^1)\xi_3,
\end{equation}
any two generators $\alpha=\alpha^i\xi_i$ and $\beta= \beta^j\xi_j$
commute if and only if $\alpha^1\beta^2-\alpha^2\beta^1=0$. The last
condition implies that $\xi_3$ must be tangent to the surface spanned by
$\alpha$ and $\beta$ if $\alpha$ and $\beta$ are independent. Such a
surface, conversely, is spanned by
$\eta_2\equiv\sin\theta\xi_1+\cos\theta\xi_2$ and $\eta_3\equiv\xi_3$,
where $\theta$ is a real parameter. Hence, we have one-parameter
($\theta$) family of distinct sets of orbits generated by a commuting
pair. This then implies that we have one-parameter degree of freedom of
relaxing a locally homogeneous manifold of Bianchi II type.

However, this fact is actually insignificant if viewing the metric as a
universal cover metric. That is, the freedom of $\theta$ can be absorbed
in the freedom of choosing the covering group $\Gamma$. Therefore we do
not have to consider the freedom of $\theta$, which can be fixed
$\theta=0$ without loss of generality. (The model based on
$\theta=\pi/2$ was presented in Ref.\cite{T} as `Type 2', which is
redundant for this reason.)
                            
Now, we have established the fact that the relaxed (universal covering)
metric can be represented by $\dh ij(x)\s i\s j$ with the basis
\reff{eq:1-nil}. The form of spacetime metric corresponding to the
two-surface orthogonal class of metrics \reff{eq:c-metric} is given by
\begin{equation}
  \label{eq:nilmetric}
  \d s^2= e^{\gamma/2}(-\d t^2+(\s1)^2)+
  R [e^P(\s3+Q\s2)^2+e^{-P}(\s2)^2],
\end{equation}
where the metric functions $\gamma$, $R$, $P$ and $Q$ are functions of
$t$ and $x$, and periodic with respect to $x$. (We have changed the
choice of $P$ and $Q$ from that in Ref.\cite{T}, since the present
choice gives us the most natural and simplest correspondence to $\bP$
and $\bQ$. See bellow.) These metric functions are related to those for
the canonical representation \reff{eq:c-metric} through
\begin{equation}
  \label{eq:n-1}
  \bar\gamma=\gamma,\; \bR=R,\; \bP=P,\; \bQ=Q-x.
\end{equation}
The reduced Einstein equations for the unbarred variables $\gamma$, $P$,
$Q$, and $R$ are therefore easily obtained by substituting these
equations into those for the canonical (i.e. barred) variables presented
in Appendix A.

Our next task is to check that we can indeed compactify the universal
cover. We choose the period for the metric functions along the $x$
axis as $2\pi$:
\begin{equation}
  \label{eq:periodNil}
  f(t,x)=f(t,x+2\pi),\quad \mbox{for }f=\gamma,\,R,\, P \mbox{ and } Q.
\end{equation}
We denote the isometry group for the spacetime metric
\reff{eq:nilmetric} as $\HIIZ$, and its subgroup which is a subgroup of
$\GII$ as $\HII\subset \GII$.  We have
\begin{equation}
  \label{eq:NI}
  \HII=\brace{\vector{2\pi n}{b}{c}\in \GII \bigg|\, n\in\Z, b,c\in\R}.
\end{equation}
The full isometry group $\HIIZ$ is generated by $\HII$ and the
$\Z_2$-isometry
\begin{equation}
  \label{eq:IIk}
  k: (x,y,z)\goes (x,-y,-z).
\end{equation}
Recall that the fundamental groups for $N_1(n)$ and $N_{-1}(n)$ are
given by Eqs.\reff{eq:Anil} with \reff{eq:pi1}. Putting
\begin{equation}
  \label{eq:II-add-1}
  g_i=\vector{2\pi n_i}{\dug i2}{\dug i3}, \quad i=1\sim3,
\end{equation}
we have to solve the relations in these fundamental groups for the
parameters $\dug ij$ and $n_i$. The solutions are easily found for
$N_1(n)$. For $N_{-1}(n)$ we have to think of $g_1$ as a composite of
$k$ and an element of $\HII$. The solutions are given by
\par\noindent
For $N_1(n)$:
\begin{eqnarray}
        \Gamma_n \wa \left\{g_1,g_2,g_3 \right\} \nonumber \\
        \wa
        \left\{
          \vector{2\pi p}{g_{1}{}^{2}}{\dug13},
          \vector{0}{0}{{2\pi\over n}(p\dug32-q\dug12)},
          \vector{2\pi q}{\dug32}{\dug33}
        \right\},
        \label{II-11}
\end{eqnarray}
where $p,q\in{\bf Z}$, $g_{i}{}^{j}\in{\bf R}$, and $p\dug32-q\dug12\neq
0$;   \par\noindent
For $N_{-1}(n)$:
\begin{eqnarray}
        \Gamma_n \wa \left\{g_1,g_2,g_3 \right\} \nonumber \\
        \wa \left\{
          k\circ\vector{2\pi p}{\dug12}{\dug13},
          \vector{0}{0}{-{2\pi p\dug32\over n}},
          \vector{0}{g_{3}{}^{2}}{g_{3}{}^{3}}
        \right\},
        \label{II-11-1}
\end{eqnarray}
where $p\in{\bf Z}$, $g_{i}{}^{j}\in{\bf R}$, and $p\dug32\neq0$. The
point here is that there exist solutions, ensuring that we can
compactify the universal cover for any closed spatial manifold
$N_{\pm1}(n)$. For completeness we remark that the parameters appearing
in $\Gamma_n$ can be fixed arbitrarily, since redefinitions of metric
functions can make the parameters equal to arbitrary values. For
example, we can take $p=\dug32=1$, and $q=\dug12=\dug13=\dug33=0$ for
$N_1(n)$, and $p=\dug32=1$, and $\dug12=\dug13=\dug33=0$ for
$N_{-1}(n)$.

Note that we did not have to impose any other condition than
Eq.\reff{eq:periodNil} for all spatial topologies $N_{\pm1}(n)$. This
results in obtaining the same reduced Einstein equations with the same
boundary conditions \reff{eq:periodNil}. Thus:
\begin{prop}
  All (two-surface orthogonal) \LUU-symmetric models of Nil type are
  dynamically equivalent.
\end{prop}
Here, the two-surface orthogonality is necessary for $N_{-1}(n)$ models
to ensure that the map $k$ is an isometry. That is, generic $N_{-1}(n)$
models with shift functions are not allowed. On the other hand, all
generic $N_1(n)$ models are dynamically equivalent.

\subsection{Case of Sol}
\label{sec:sol}

The invariant one-forms are given by
\begin{equation}
  \label{eq:1-sol}
  \s1=\d x,\;
  \s2=\rcp{\sqrt2}(e^{\q x}\d y+e^{-\q x}\d z),\; 
  \s3=\rcp{\sqrt2}(-e^{\q x}\d y+e^{-\q x}\d z),
\end{equation}
where $\q>0$ is a positive parameter introduced for convenience.

Using these 1-forms we can write the two-surface orthogonal metric as
\begin{equation}
  \label{eq:sol-metric}
  \d s^2= e^{\gamma/2}(-\d t^2+(\s1)^2)+
    R [e^P(\s2{}'+Q\s3{}')^2+e^{-P}(\s3{}')^2],
\end{equation}
where $\gamma$, $P$, $Q$, and $R$, are functions of $t$ and $x$, and
they are assumed to be periodic in $x$. We have defined
\begin{equation}
  \label{eq:s-1}
  \s2{}'\equiv \rcp{\sqrt2}(\s2-\s3)=e^{\q x}\d y,\quad
  \s3{}'\equiv \rcp{\sqrt2}(\s2+\s3)=e^{-\q x}\d z.
\end{equation}
(The choice of metric functions $P$ and $Q$ here is different from that
in Ref.\cite{T}. We have done so for apparent simplicity, but we should
keep in mind that at the homogeneous limit the metric is not
diagonalizable with respect to $(\s1,\s2{}',\s3{}')$, in contrast to
with respect to $(\s1,\s2,\s3)$, for vacuum spacetimes.)
The metric functions
defined with Eq.\reff{eq:sol-metric} are related to those in the canonical
representation \reff{eq:c-metric} through
\begin{equation}
  \label{eq:s-10}
  \bar\gamma=\gamma,\; \bR=R,\; 
  \bP=P+2\q x,\; 
  \bQ=e^{-2\q x}Q.
\end{equation}
The reduced Einstein equations for $\gamma$, $P$, $Q$, and $R$ can be
obtained by substituting these equations into those presented in Appendix
\ref{sec:gc}.

For compactifications, see Appendix \ref{sec:comsol}. In particular, the
isometry group of the metric \reff{eq:sol-metric} is given by the
$\HsixZ$ presented in the Appendix. Therefore we can compactify the
universal cover possessing this metric for all $A$. If we choose the
period along $x$-axis for the metric functions as $2\pi$, we must put
$c_3=2\pi$. Accordingly, we must choose the parameter $\q$ so that
$e^{2\pi \q}$ equals to the greater absolute eigenvalue of the matrix
$A$, since we have assumed $\q>0$. Note that the characteristic
polynomial \reff{eq:apsol-9} depends only on $\Tr A$.  Let $n\equiv\Tr
A$. Then we obtain
\begin{equation}
  \label{eq:sol-q}
  \q=\rcp{2\pi}\log{\abs{n}+\sqrt{n^2-4}\over 2}.
\end{equation}
The reduced Einstein equations for the (unbarred) metric functions
depends (only) on the parameter $\q$, and the boundary conditions are the
common simple periodic boundary condition. Moreover, $\q$ depends only
on $\abs n$. We thus conclude:
\begin{prop}
  An \LUU-symmetric model of Sol type is specified with an element
  $A\in\omcg{T^2}\simeq\SL{2,\Z}$ such that $\abs{\Tr A}>2$. Let $A_1$
  and $A_2$ be such matrices. Then, the corresponding models are
  dynamically equivalent if $\abs{\Tr A_1}=\abs{\Tr A_2}$.
\end{prop}
From this fact, we find it suffices to consider the models with the
one-parameter family $S(n)$ ($n>2$), defined in Eq.\reff{eq:Asol}, as
representatives.

\subsection{Case of $E^3$}
\label{sec:e3}

The corresponding Bianchi types are Bianchi I and \VII0.  More
precisely, geometry $E^3$ is the Euclid space $\R^3$ with the standard
metric $\d x^2+\d y^2+\d z^2$. The isometry group of $E^3$ is therefore
formed by the translations $\R^3$ and rotations $\O3$, and is isomorphic
to the \Poin\ group, $\mathrm{Isom}E^3\simeq\IO3$. $\IO3$ contains two
simply transitive subgroups, the Bianchi I ($\GI\simeq\R^3$) and \VII0
($\Gseven$) groups, so that the correspondence above is arrived at.

We can make the inhomogeneous metrics from both $\GI$ and
$\Gseven$. However, the effects are not equivalent. First, consider the
metric made from $\GI$.  Since the invariant one-forms coincide with the
usual coordinate basis $\d x$, $\d y$, and $\d z$, the spacetime metric
and the boundary conditions are the same as the canonical ones
\reff{eq:g1metric} and \reff{eq:g1bc}. We find in particular that the
metric functions for $E_1$ and $E_2$ are simply periodic.  For $E_3$,
$E_4$, and $E_6$, however, the metric functions must obey the last
condition in Eqs.\reff{eq:g1bc}, which does not imply the simple
periodic conditions, (though they have period $6\pi$, $4\pi$, and $6\pi$
for $E_3$, $E_4$, and $E_6$, respectively).  As long as we adhere to
the metric obtained from $\GI$, this feature is inevitable, since the
closed manifolds corresponding to $E_3$, $E_4$, and $E_6$ cannot be
realized in the Bianchi I group $\GI$, or $\GIZ$. Here, $\GIZ$ is
generated by $\GI$ and the $\Z_2$-map $h:\; (x,y,z)\goes (x,-y,-z)$,
which is an isometry of the metric \reff{eq:g1metric}. Precisely, the
fundamental group $\pi_1(E_1)$ corresponding to $E_1$ can be embedded in
$\GI$, and $\pi_1(E_2)$ can be embedded in the $\GIZ$. However, other
fundamental groups can be embedded neither in $\GI$ nor $\GIZ$. Since the
isometry group of the metric \reff{eq:g1metric} becomes a subgroup of
$\GIZ$, provided that the metric functions are all periodic with period
$2\pi$, this implies that the embeddings of $\pi_1$ for $E_3$, $E_4$,
and $E_6$ into the subgroup are impossible. Thus, the appropriate
boundary conditions cannot be the simple periodic conditions for them.

On the other hand, all the fundamental groups $\pi_1(E_\lambda)$
$(\lambda=1,2,3,4,6)$ can be embedded into $\Gseven$ \cite{KTH}, which
fact implies that the boundary conditions can become simple periodic
ones for the metric obtained from $\Gseven$ for all $E_\lambda$. We
present an explicit prescription bellow.

The invariant one-forms for Bianchi \VII0 are given by
\begin{equation}
  \label{eq:e3-1}
  \s1=\d x,\;
  \s2=\cos\q x\,\d y+\sin\q x\,\d z,\; 
  \s3=-\sin\q x\,\d y+\cos\q x\,\d z,
\end{equation}
where $\q>0$ is a positive parameter.
Using these 1-forms we can write the two-surface orthogonal metric as
\begin{equation}
  \label{eq:e3-2}
  \d s^2= e^{\gamma/2}(-\d t^2+(\s1)^2)+
    R [e^P(\s2+Q\s3)^2+e^{-P}(\s3)^2],
\end{equation}
where $\gamma$, $P$, $Q$, and $R$, are functions of $t$ and $x$, and
they are assumed to be periodic in $x$ with period $2\pi$. The parameter 
$\q$ will be chosen so that we can compactify the universal cover. (The
role of $\q$ is similar to the $\q$ appearing in the Sol model.)
The correspondence to the barred metric functions in the canonical
metric \reff{eq:c-metric} is given by
\begin{eqnarray}
  e^{\bP(x)} \wa \Phi(P(x),Q(x),\cos\q x,\sin\q x), \quad
  \bQ(x) = \Psi(P(x),Q(x),\cos\q x,\sin\q x), \nonumber \\
  e^{P(x)} \wa \Phi(\bP(x),\bQ(x),\cos\q x,-\sin\q x), \quad
  Q(x) = \Psi(\bP(x),\bQ(x),\cos\q x,-\sin\q x).
  \label{eq:e3-2.2}
\end{eqnarray}
where the two functions $\Phi$ and $\Psi$ are defined by, for arbitrary
$P$, $Q$, $c$, and $s$,
\begin{eqnarray}
  \Phi(P,Q,c,s) \wa e^P(c-s Q)^2+e^{-P}s^2, \nonumber \\
  \Psi(P,Q,c,s) \wa 
  \frac{e^P(c-s Q)(s+c Q)-e^{-P}c s}{\Phi(P,Q,c,s)}.
  \label{eq:e3-2.2.5}
\end{eqnarray}
Note that the vacuum Einstein equations in terms of the unbarred
variables contain the parameter $\q$.

As usual, we represent the elements of the Bianchi group in the column
vector form. The multiplication rule for $\Gseven$ is given by
\begin{equation}
  \label{eq:e3-3}
  {\vector abc}_\q{\vector xyz}_\q={\svector{a+x}{\svector bc+R_{\q
      a}\svector yz}}_\q, \;
  {\vector abc}_\q\inv={\svector{-a}{-R_{-\q a}\svector bc}}_\q,
\end{equation}
where $R_{\q a}$ is the rotation matrix by angle $\q a$;
\begin{equation}
  \label{eq:e3-3.5}
  R_{\q a}=\smatrix{\cos\q a}{-\sin\q a}{\sin\q a}{\cos\q a}.
\end{equation}
The subscripts $\q$ appearing in the column vectors are to remind that
the multiplication rule is defined with respect to $\q$. The one-forms
\reff{eq:e3-1} are invariant under this action.

We denote the isometry group of the metric \reff{eq:e3-2} which is a
subgroup of $\Gseven$ as $\Hseven$. It is given by
\begin{equation}
  \label{eq:e3-4}
  \Hseven=
  \brace{{\vector{2\pi n}{b}{c}}_\q\in\Gseven\bigg|\, n\in\Z,\; b,c\in\R}.
\end{equation}
The (full) isometry group $\HsevenZ$ of the metric \reff{eq:e3-2} is
generated by $\Hseven$ and the $\Z_2$-isometry $h:\; (x,y,z)\goes
(x,-y,-z)$.

We can perform the embedding of every $\pi_1(E_\lambda)$ both with and
without the $\Z_2$-isometry $h$, as we can read from Sec.V A of
Ref.\cite{KTH}. (But, be careful with the differences in the choice of
representations of the relations in $\pi_1$.) When we do not use $h$,
i.e., embed the fundamental group $\pi_1(E_\lambda)$ into $\Hseven$, the
appropriate value of $\q$ is simply given by
\begin{equation}
  \label{eq:e3-5}
  \q=\rcp\lambda.
\end{equation}
Hence all the models seem to be dynamically inequivalent. However, we
can choose $\q=1$ for $\lambda=2$, and $\q=1/3$ for $\lambda=6$, if we
use $h$. For example, in the case of $E_6$ one can confirm that the
solution of the embedding into $\Hseven$ is given by
\begin{eqnarray}
        \Gamma_6 \wa \left\{g_1,g_2,g_3 \right\} \nonumber \\
        \wa
        \left\{
          {\vector{2\pi}{\dug12}{\dug13}}_{\rcp6},
          {\vector{0}{\dug22}{\dug23}}_{\rcp6},
          {\svector{0}{-R_{\pm \pi/3}\svector{\dug22}{\dug23}}}_{\rcp6}
        \right\}
        \label{e3-6}
\end{eqnarray}
with $\q=1/6$. Here, $\dug ij$ are real parameters.
Also, we can embed the fundamental group into $\HsevenZ$ as
\begin{equation}
        \Gamma_{6'} =
        \left\{
          h\circ{\vector{2\pi}{\dug12}{\dug13}}_{\rcp3},
          {\vector{0}{\dug22}{\dug23}}_{\rcp3},
          {\svector{0}{R_{2\pi/3}\svector{\dug22}{\dug23}}}_{\rcp3}
        \right\}
        \label{e3-7}
\end{equation}
with $\q=1/3$. Since the fundamental group of $E_3$ can be embedded into 
$\Hseven$ with $\q=3$ as
\begin{equation}
        \Gamma_3 =
        \left\{
          {\vector{2\pi}{\dug12}{\dug13}}_{\rcp3},
          {\vector{0}{\dug22}{\dug23}}_{\rcp3},
          {\svector{0}{-R_{\pm 2\pi/3}\svector{\dug22}{\dug23}}}_{\rcp3}
        \right\},
        \label{e3-8}
\end{equation}
the $E_6$ and $E_3$ models are dynamically equivalent.
(The significance of the parameters $\dug ij$ is the same as in the case 
of Nil. In particular, we can take $\dug12=\dug13=0$, $\dug22=1$,
and $\dug23=0$, in all cases above.)

As a result, we have only three dynamically inequivalent classes;
\begin{prop}
  The $E_1$, $E_3$, and $E_4$ models comprise a set of representatives
  for the dynamical equivalence in the two-surface orthogonal
  \LUU-symmetric models of $E^3$ type.
\end{prop}
For the generic models of $E^3$ type with the shift functions, all the
$E_\lambda$ models are dynamically inequivalent, since the map $h$ is
not an isometry.

To end this subsection, we remark again that the \LUUs\ models of $E^3$
type best match with Bianchi \VII0, not Bianchi I. However, if limited
to the $T^3$ (or $T^3/\Z_2$) model, which corresponds to $E_1$ ($E_2$),
what it naturally corresponds is Bianchi I. To make later discussions
easier, it may therefore be convenient to put forward the following
convention.

\conventionmark When speaking of the ``$E^3$ models'', we understand
them to have correspondence to Bianchi \VII0. In particular, the
``unbarred variables'' for an $E^3$ model are those in the metric
\reff{eq:e3-2}. However, when speaking of the ``$T^3$ model'', we
understand it to have correspondence to Bianchi I. Therefore the
unbarred variables coincide with barred variables appearing in the
canonical metric \reff{eq:c-metric}.

\section{Translation and Reflection symmetries}
\label{sec:sym}

The Gowdy equations \reff{eq:a-2} have several symmetries. First of all,
from the fact that these equations do not have explicit $x$-dependence,
they have the natural translation symmetry with respect to $x\goes x-a$,
where $a$ is a real parameter. This means that if $(\bP(t,x),\bQ(t,x))$
is a solution for the equations, then $(\bP(t,x+a),\bQ(t,x+a))$ is also
a solution. One more symmetry that naturally comes to our attention is
the reflection symmetry with respect to $x\goes -x$, which is a result
of the invariance of the equations under the transformation $\vbrow
x\goes - \vbrow x$. This symmetry means that if $(\bP(t,x),\bQ(t,x))$ is
a solution for the equations, then $(\bP(t,-x),\bQ(t,-x))$ is also a
solution. Another feature that results from these symmetries is that (in
the Hamiltonian picture) an initial data that is symmetric with respect
to $x\goes x+a$ or $x\goes -x$ maintains the same symmetry under the
time evolution. This point will be discussed in the second subsection
bellow.

For the usual Gowdy $T^3$ model, both $x\goes x-a$ and $x\goes -x$
transformations acting on $(\bP(t,x),\bQ(t,x))$ preserve the boundary
conditions on $\bP$ and $\bQ$, so that they express true symmetries. For
Nil and Sol models, however, these transformations do not preserve the
boundary conditions, so that they do not express true
symmetries. Nevertheless, we can find translation symmetries for Nil and
Sol models in a generalized sense, and also find (generalized)
reflection symmetries for them. We also find that the $T^3$ model admits
larger amounts of reflection symmetries than mentioned above. In the
first subsection bellow, we obtain these translation and reflection
symmetries in a systematic way. Based on this result, in the second
subsection we discuss the dynamical point of view of the symmetries,
which reveals a manifestation of the influence of spatial topologies to
dynamics.

\subsection{Translation and Reflection Transformations}

All symmetries concerning the Gowdy equations \reff{eq:a-2} should have
their origin in the metric \reff{eq:c-metric}. For example, this metric
is invariant under the \diffeo\ $x\goes x-a$, {\it up to} the
redefinition of the metric functions
$(\bP(x),\bQ(x))\goes(\bP(x+a),\bQ(x+a))$ (with the similar one for
$\blambda$). (For simplicity, we omit writing the argument $t$ in the
metric functions.) This redefinition accounts for the translation
symmetry mentioned above. Recall that, if a metric is a solution for the
vacuum Einstein equation, then so is the metric induced from a (spatial)
\diffeo. Thus, in general, if such a \diffeo\ leaves the metric
\reff{eq:c-metric} invariant up to a redefinition of the metric
functions, this redefinition defines a symmetry for the Gowdy equations.

Consider a \diffeo\ $\phi:\; (x,y,z)\goes
(\phi_x(x,y,z),\phi_y(x,y,z),\phi_z(x,y,z))$ with $t$ fixed. As
remarked, $\phi$ must preserve the characteristic form of the metric
\reff{eq:c-metric} so that a redefinition of the metric functions can
make the metric invariant. In particular, to preserve the isothermal
form $e^{\blambda/2}(-\d t^2+\d x^2)$ of the metric we must have
$\phi_x(x,y,z)=\pm x-a$. Moreover, $(y,z)\goes (\phi_y,\phi_z)$ must be
a linear transformation on each $y$-$z$ plane. That is, we must have
\begin{equation}
  \label{eq:trt-1}
\phi:\; (x,{\bf y})\goes (\pm x-a,L{\bf y}),
\end{equation}
where $L\in \SLZ$ is a constant matrix and ${\bf y}$ is the column form
vector of $(y,z)$. Here, $\SLZ\equiv \{ m\in \GL{2,\R}| \det(m)=\pm1\}$.
(We do not have to consider the translations in the $y$-$z$ plane of the
type $(x,y,z)\goes (x,y+y_0,z+z_0)$, which are isometries for the metric
\reff{eq:c-metric}, since these give rise to no effect to the Gowdy
equations.)  We decompose the $\phi_x$ part into $x\goes x-a$ and
$x\goes -x$, and will call symmetries involved with the former {\it
  (generalized) translation symmetries} and call symmetries involved
with the latter {\it (generalized) reflection symmetries}. We can also
consider symmetries generated by \diffeos\ involved with the identity
$x\goes x$, but since this type of symmetry is of little interest from
the dynamical point of view discussed in the next subsection, we do not
consider them here.  We can find all possible translation and reflection
symmetries defined above by finding $\phi_y(y,z)$ and $\phi_z(y,z)$ (or
$L$) for each case, as follows.

We first consider the translation symmetries. As we will see in the next
subsection, this symmetry implies that (in the Hamiltonian picture) a
symmetric initial data with respect to a (generalized) translation
maintains its symmetry under the time evolution. At the (locally)
homogeneous limit, this implies that the \diffeo\ that gives rise to
this symmetry should coincide with an isometry for the metric. Recall
that there is a three-dimensional isometry group $G$ for each case of
$T^3$, $E^3$, Nil, and Sol at the homogeneous limit.  It contains a
two-dimensional commutative isometry subgroup $H\simeq \R^2$ that is
preserved as isometry group when the metric becomes \LUUs\ but
inhomogeneous. Then, there is a one-parameter isometry subgroup $K$
which acts transversely to $H$. In the \LUUs\ but inhomogeneous cases,
the actions of $K$ are served as the desired translations. As easily
found from the multiplication rules for the Bianchi I, \VII0, II, and
\VI0 groups, they are, respectively, given by
\begin{eqnarray}
  \label{eq:sym-1-1-1}
  T_a &:& (x,y,z)\goes (x-a,\, y,\, z), \\
  \label{eq:sym-1-1-2}
  E_a &:& (x,y,z)\goes 
  (x-a,\, \cos\q a\, y+\sin\q a\, z,\, -\sin\q a\, y+\cos\q a\, z), \\
  \label{eq:sym-1-2}
  N_a &:& (x,y,z)\goes (x-a,\, y,\, z-ay), \\
  \label{eq:sym-1-3}
  S_a &:& (x,y,z)\goes (x-a,\, e^{2qa}y,\, e^{-2qa}z).
\end{eqnarray}
Since these leave the invariant one-forms for, respectively, Bianchi I,
\VII0, II, and VI${}_0$ types invariant, the corresponding transformations
(redefinitions) takes the simplest form for the unbarred metric
functions:
\begin{eqnarray}
  \label{eq:sym-2-0}
  T_{a*} &:& (\bP(x),\bQ(x))\goes (\bP(x+a),\bQ(x+a))
  \quad\mbox{for metric functions in \reff{eq:c-metric}},\\
  \label{eq:sym-2-1}
  E_{a*} &:& (P(x),Q(x))\goes (P(x+a),Q(x+a))
  \quad\mbox{for metric functions in \reff{eq:e3-2}},\\
  \label{eq:sym-2-2}
  N_{a*} &:& (P(x),Q(x))\goes (P(x+a),Q(x+a))
  \quad\mbox{for metric functions in \reff{eq:nilmetric}},\\
  \label{eq:sym-2-3}
  S_{a*} &:& (P(x),Q(x))\goes (P(x+a),Q(x+a))
  \quad\mbox{for metric functions in \reff{eq:sol-metric}}.
\end{eqnarray}
Here, we assume $a=2\pi/n$ for a positive integer $n$.  Using the
relations \reff{eq:e3-2.2}, \reff{eq:n-1} and \reff{eq:s-10}, $E_{a*}$,
$N_{a*}$ and $S_{a*}$ expressed with the canonical metric
\reff{eq:c-metric} are found to be
\begin{eqnarray}
  \label{eq:sym-3}
  E_{a*} &:& (\bP(x),\bQ(x))\goes
  (\,\log \Phi(\bP(x+a),\bQ(x+a),\cos\q a,-\sin\q a),
 \nonumber \\
  && \hspace{8em}
  \Psi(\bP(x+a),\bQ(x+a),\cos\q a,-\sin\q a)\,), \\
  N_{a*} &:& (\bP(x),\bQ(x))\goes (\bP(x+a),\bQ(x+a)+a), \\
  S_{a*} &:& (\bP(x),\bQ(x))\goes (\bP(x+a)-2\q a,e^{2\q a}\bQ(x+a)).
\end{eqnarray}
The functions $\Phi$ and $\Psi$ are defined in Eq.\reff{eq:e3-2.2.5}.
$T_{a*}$, $E_{a*}$, $N_{a*}$ and $S_{a*}$ preserve the boundary
conditions for the \LUUs\ models of, respectively, $T^3$, $E^3$, Nil and
Sol types, so they define the translation symmetry for each type. Note
that the translations for each type form the infinite cyclic group $\Z$
for a fixed $a$, or form the cyclic group $\Z_n$ for a fixed $a=2\pi/n$
if the (periodic) boundary conditions are taken into account.

Next, we consider the reflection symmetries, by which we mean the
symmetries generated by the reflection transformations defined as
follows.

\nd \defmark Consider the map $R:\; (x,{\bf y})\goes (-x, L{\bf y})$ for
a given $L\in\SLZ$.  The symmetry transformation $R_*$ induced from $R$,
acting on the variables $(P,Q)$ of a given model, is called a {\it
  reflection transformation}, if the following three conditions are
fulfilled;
\begin{description}
\item[(i)] For $R$ (and $R_*$) to form the $\Z_2$-group together with
  the identity $\mathrm{id}$,
\begin{equation}
  \label{eq:sym-3.5}
  R^2\equiv R\circ R=\mathrm{id}.
\end{equation}
\item[(ii)]
As explained later,
\begin{equation}
  \label{eq:sym-6}
  \Conj(R)T\equiv R\circ T\circ R\inv=T\inv,
\end{equation}
where $T$ is the map inducing the translation transformation of the
given model. (Hence, $T$ is one of Eqs.\reff{eq:sym-1-1-1} $\sim$
\reff{eq:sym-1-3}.)
\item[(iii)] $R_*$ preserves the boundary conditions. That is, if
  variables $(P,Q)$ satisfy the boundary conditions appropriate for the
  model, $R_*(P,Q)$ also satisfy the same boundary conditions.
\end{description}

First, from condition (i), $L$ must be one of
\begin{equation}
  \label{eq:sym-4}
  L_0\equiv\smatrix 1001,\quad 
  L_\pi\equiv\smatrix{-1}00{-1}, \mbox{ or}\quad 
  L_{l_\theta}\equiv
  \smatrix{\cos 2\theta}{\sin 2\theta}{\sin 2\theta}{-\cos 2\theta},
\end{equation}
where $\theta\in [0,\pi)$ is a real parameter. These act on the $y$-$z$
plane as, respectively, the identity, the rotation by angle $\pi$, and
the reflection with respect to the line $l_\theta$ which passes through
the origin with angle $\theta$. Thus we have obtained the following
candidates for the reflection $R$:
\begin{eqnarray}
  \label{eq:sym-5}
  R_0 &:& (x,y,z)\goes (-x,y,z), \\
  R_\pi &:& (x,y,z)\goes (-x,-y,-z), \\
  R_{l_\theta} &:& (x,y,z)\goes 
  (-x,y\cos 2\theta+z\sin 2\theta,y\sin 2\theta-z\cos 2\theta).
\end{eqnarray}

Condition (ii) concerns a ``compatibility'' with the natural translation
obtained earlier. That is, the reflection about a generic point $x=a$
should be induced from that about $x=0$ by a translation. We can
equivalently require that the conjugation, $\Conj(R)T$, of the
translation $T$ by a reflection $R$ reverse the original translation, as
indicated by Eq.\reff{eq:sym-6}. For example, the conjugation of the
translation $T_a$ for the $T^3$ model by the map $R_0$ is given by $
\Conj(R_0)T_a = R_0\circ T_a\circ (R_0)\inv= R_0\circ T_a\circ R_0=
T_{-a}$. This shows $R_0$ is compatible with $T_a$. On the other hand,
say, the conjugation of $N_a$ by $R_0$ gives $ \Conj(R_0)N_a=R_0\circ
N_a\circ R_0 = T_{-2a}\circ N_a$, which is not the reverse of $N_a$. All
the compatible relations are given as follows:
\begin{eqnarray}
  \label{eq:sym-7}
  \mbox{For $T^3$ model:}\hspace{3em}
  \Conj(R_0\,)T_a \wa T_{-a}, \\
  \Conj(R_\pi\,)T_a \wa T_{-a}, \\
  \Conj(R_{l_\theta})T_a \wa T_{-a} \quad (\theta\in [0,\pi)),\\
  \mbox{For $E^3$ model:}\hspace{3em}
  \Conj(R_{l_\theta})E_a \wa E_{-a} \quad (\theta\in [0,\pi)),\\
  \mbox{For Nil model:}\hspace{3em}
  \Conj(R_{l_\theta})N_a \wa N_{-a} \quad (\theta=0,{\pi\over2}),\\
  \mbox{For Sol model:}\hspace{3em}
  \Conj(R_{l_\theta})S_a \wa S_{-a} \quad (\theta={\pi\over4},{3\over4}\pi).
\end{eqnarray}

The transformations for the metric functions induced from these maps are
given as follows:
\def\c{c}
\def\s{s}
\begin{eqnarray}
  \label{eq:sym-8-1-1}
  R_{0*},\,R_{\pi*} &:& (\bP(x),\bQ(x))\goes (\bP(-x),\bQ(-x)), \\
  R_{l_\theta *} &:& (\bP(x),\bQ(x))\goes
  (\,\log\Phi(\bP(-x),\bQ(-x),-\sin 2\theta,\cos 2\theta),
  \nonumber \\
  & & \hspace{8em}
  -\Psi(\bP(-x),\bQ(-x),-\sin 2\theta,\cos 2\theta)\, ),
  \label{eq:sym-8-1-2}
\end{eqnarray}
where
the functions $\Phi$ and $\Psi$ are defined in Eq.\reff{eq:e3-2.2.5}.
The explicit forms for Nil ($\theta=0,\pi/2$) and Sol
($\theta=\pi/4,3\pi/4$) models are
\begin{eqnarray}
  \label{eq:sym-10-1}
  R_{l_0*},\,R_{l_{\pi/2}*} &:& (\bP(x),\bQ(x))\goes
  (\bP(-x),-\bQ(-x)), \\
  \label{eq:sym-10-2}
  R_{l_{\pi/4}*},\,R_{l_{3\pi/4}*} &:& (\bP(x),\bQ(x))\goes
  \paren{ \bP+\log(\bQ^2+e^{-2\bP}),\;
  \frac{\bQ}{\bQ^2+e^{-2\bP}} }\bigg|_\shortPQ.
\end{eqnarray}
Note that $R_{0*}=R_{\pi*}$, $R_{l_0*}=R_{l_{\pi/2}*}$, and
$R_{l_{\pi/4}*}=R_{l_{3\pi/4}*}$.

As a final task we are left with checking condition (iii). For the
$T^3$ model, we at once see that if $\bP(x)$ and $\bQ(x)$ are periodic
functions, then their images by $R_{0*}$ and $R_{l_\theta*}$ are also
periodic functions. Thus, $R_{0*}$ and $R_{l_\theta*}$ give reflections
for the $T^3$ model. For the $E^3$, Nil and Sol models it is convenient
to work with the unbarred variables. Interestingly the transformations
\reff{eq:sym-8-1-2} (for $E^3$), \reff{eq:sym-10-1} (for Nil) and
\reff{eq:sym-10-2} (for Sol) are written with the unbarred variables in
exactly the same form (except for the disappearance of bars);
\par\noindent
For $E^3$:
\begin{eqnarray}
  R_{l_\theta *} &:& (P(x),Q(x))\goes
  (\,\log\Phi(P(-x),Q(-x),-\sin 2\theta,\cos 2\theta),
  \nonumber \\
  & & \hspace{8em}
  -\Psi(P(-x),Q(-x),-\sin 2\theta,\cos 2\theta)\, ),
  \label{eq:sym-11-0}
\end{eqnarray}
\par\noindent
For Nil:
\begin{equation}
  \label{eq:sym-11-1}
  R_{l_0*} :\; (P(x),Q(x))\goes
  (P(-x),-Q(-x)),
\end{equation}
For Sol:
\begin{equation}
  \label{eq:sym-11-2}
  R_{l_{\pi/4}*} :\; (P(x),Q(x))\goes
  \paren{ P+\log(Q^2+e^{-2 P}),\;
  \frac{Q}{Q^2+e^{-2 P}} }\bigg|_\shortPQ.
\end{equation}
From these it is trivial to see that the periodicity of the unbarred
variables are preserved for $E^3$, Nil and Sol. This confirms that these
transformations do give reflections for the corresponding models. Table
\ref{tab:sym} bellow summarizes the translation and reflection
transformations for each model.
\begin{table}[hbtp]
  \begin{center}
\begin{tabular}[c]{|c|c|c|} \hline
  Model & Translation & \hspace*{.5em}Reflections\hspace*{.5em} \\ \hline
  $T^3$ & $T_{a*}$ & $R_{0*}, \;R_{l_\theta *}$ 
   \\ \hline
  $E^3$ & $E_{a*}$ & $R_{l_\theta *}$ 
   \\ \hline
  Nil & $N_{a*}$ &  $R_{l_0*}$  \\ \hline
  Sol & $S_{a*}$ &  $R_{l_{\pi/4}*}$  \\ \hline
\end{tabular}
    \caption{\small The translation and reflection
      symmetry transformations. For reflections of $T^3$ and $E^3$,
      $\theta\in [0,\pi)$. }
    \label{tab:sym}
  \end{center}
\end{table}

Remark that we have inclusion relations. The $T^3$ model has the largest
set of reflections. The $E^3$ models have part of that for the $T^3$
model. The Nil and Sol models have parts of that for the $E^3$ models.

\subsection{Dynamical Interpretation}

\def\P{{\cal P}}
\def\O{{\cal O}}
\def\Oc{\O_{\rm com}}
\def\S{{\cal S}}

As well known \cite{Gow2}, the Gowdy equations \reff{eq:a-2} admit a
Hamiltonian formulation. Let $\pi_P(x)$ and $\pi_Q(x)$ be, respectively,
conjugate momenta of $P(x)$ and $Q(x)$. The phase space $\P$ is spanned
by the four functions $(P(x),Q(x),\pi_P(x),\pi_Q(x))$ (or similarly for
variables with bars). Consider the set $\O$ of maps on $\P$,
$\O\equiv\{o| o: \P\goes \P\}$. The Hamiltonian flow $\psi_\tau$, which
generates the time evolution by time $\tau$, is naturally regarded as an
element in $O$, $\psi_\tau\in \O$, for a fixed $\tau$. $\psi_0$ is
supposed to be the identity.

What we are interested in here are maps (operators) which commute with
$\psi_\tau$, so we define
\begin{equation}
  \label{eq:h-1}
  \Oc\equiv \{f\in\O |\; {}^\forall\tau\in\R,\quad
  f\circ \psi_\tau= \psi_\tau\circ f \,\}.
\end{equation}
We call an $f\in \Oc$ a {\it symmetry operator}. (If the operator $f$
was a smooth flow like $\psi_\tau$ we would be able to reformulate this
commutativity to the vanishing of a Poisson bracket. However, since we
consider discrete operators as $f$ this is not the case. As a result, we
do not actually need symplectic structures if only the flow $\psi_\tau$
is properly defined.)  For an initial data $\gamma\in\P$,
$\psi_\tau(\gamma)$ is a solution for the Einstein equation as a
function of $\tau$. If there is a symmetry operator $f$, then the
function of $\tau$, $f(\psi_\tau(\gamma))$, is also a solution, since
this equals to $\psi_\tau(f(\gamma))$, which is the solution with the
initial data $f(\gamma)$. Therefore a symmetry operator generates a (in
general, distinct) solution from a solution.

Now, note that, while a symmetry transformation obtained in the previous
subsection is defined in the configuration space spanned by
$P$ and $Q$, it induces an operator on the phase space $\P$ by
differentiating the configuration variables with respect to $\tau$. We
use in particular the Hamiltonian equations
\begin{equation}
\label{eq:h-1.5}
\dot P=\pi_P,\quad \dot Q=e^{-2P}\pi_Q.
\end{equation}
(This relation is the same for all sets of barred or unbarred
variables.)  Thus, the symmetry transformations are also regarded as an
operator on $\P$. This operator is clearly a symmetry operator.
Conversely, if there is a symmetry operator, it gives a symmetry
transformation, as seen from the previous paragraph. Thus, in effect,
symmetry transformations and operators are equal entities. Note that,
since we have exhausted all the symmetry transformations concerning
translation and reflection, we now know all the symmetry operators for
them.

To represent the symmetry operator obtained from a symmetry
transformation we use the same symbol as that of the symmetry
transformation. For example, the translation transformation $T_{a*}$ for
$T^3$ model defined in Eq.\reff{eq:sym-2-0} naturally gives the
operator, denoted also as $T_{a*}$, that simply shifts the argument by a
constant:
\begin{equation}
  \label{eq:h-a1}
  T_{a*}: (\bP(x),\bQ(x),\bar\pi_P(x),\bar\pi_Q(x)) \goes
  (\bP(x+a),\bQ(x+a),\bar\pi_P(x+a),\bar\pi_Q(x+a)).
\end{equation}
The translation symmetry operators $E_{a*}$ (for $E^3$), $N_{a*}$ (for
Nil) and $S_{a*}$ (for Sol) are also defined by similar maps if the
phase space variables are taken as the ``unbarred''
ones. (cf. Eqs.\reff{eq:sym-2-1}, \reff{eq:sym-2-2}, and
\reff{eq:sym-2-3}.)

An interesting consequence of the symmetry operators comes from
considering data which are invariant with respect to this
operator. For $f\in\Oc$, let
\begin{equation}
  \label{eq:h-2}
  \S(f)\equiv \{\gamma\in\P | f(\gamma)=\gamma \}.
\end{equation}
For example, if $f$ is a translation symmetry operator $T_{2\pi/n*}$,
the set $\S(T_{2\pi/n*})$ is comprised of data in $\P$ which are
translation-symmetric with period $2\pi/n$. If $f$ is a reflection
symmetry operator $R_*$, the set $\S(R_*)$ is comprised of
reflection-symmetric data in $\P$. The remarkable fact is that {\it
  these symmetries are preserved dynamically}, since Eq.\reff{eq:h-1},
together with $f(\gamma)=\gamma$, implies
$f(\psi_\tau(\gamma))=\psi_\tau(\gamma)$ for all $\tau$. In other words,
any data $\gamma\in \S(f)$ remains within $\S(f)$ under the dynamical
flow, i.e., $\psi_\tau(\S(f))\subseteq\S(f)$. Thus, $\S(f)$ defines an
{\it invariant subset} of $\P$ for the flow $\psi_\tau$.

We interpret the invariant subset $\S(f)$ as a dynamical character of
the model. Motivated by this interpretation, we compare $\S(f)$ to see
if there are distinctions in dynamical characters of the models. To
this, recall that if we think of the phase space $\P$ as being spanned
by the unbarred variables, $\P$ for every model is a space of four
functions which are all periodic functions. (We think that the barred
and unbarred variables coincide for the $T^3$ model.)  Therefore the
phase spaces for the four models are naturally identified. In this view,
the flow $\psi_\tau$ depends on the type of the model. Strictly
speaking, however, since the ``constraints'' equations corresponding to
Eq.\reff{eq:a-5} take different forms for the $E^3$, Nil and Sol models,
this identification may be justified only approximately, but we neglect
this point.

First, consider the translations. As already mentioned the translation
operator for every model is obtained by simply shifting the spatial
coordinate like Eq.\reff{eq:h-a1} for the unbarred variables. Hence, all
the invariant subsets $\S(T_{a*})$, $\S(E_{a*})$, $\S(N_{a*})$, and
$\S(S_{a*})$ for $a=2\pi/n$, where $n$ is a positive integer, are
spanned by periodic data with period $2\pi/n$. In this sense, we
roughly represent
\begin{equation}
  \label{eq:d-0}
  \S(T_{a*})\simeq \S(E_{a*})\simeq \S(N_{a*})\simeq \S(S_{a*}).
\end{equation}
Therefore we conclude that for the property of preservation of
translation symmetry, there is no significant distinction between the
four models.

Our main interest is therefore in the reflections. As we remarked,
reflection operator
\begin{equation}
  \label{eq:d-1}
  R_{0*}: \; (\bP(x),\bQ(x),\bar\pi_P(x),\bar\pi_Q(x)) \goes
  (\bP(-x),\bQ(-x),\bar\pi_P(-x),\bar\pi_Q(-x))
\end{equation}
does not preserve the boundary conditions for $E^3$, Nil and Sol models,
so that it is not relevant for these three models. At first sight, the
similar operator acting on unbarred variables
\begin{equation}
  \label{eq:d-2}
  R_{0*}': \; (P(x),Q(x),\pi_P(x),\pi_Q(x)) \goes
  (P(-x),Q(-x),\pi_P(-x),\pi_Q(-x)),
\end{equation}
which does preserve the boundary conditions for the three models, seems
to be a natural reflection operator for them. Indeed, we can regard the
spatial coordinate $x$ for the unbarred variables as a natural
coordinate in view of the translation property shown above. That is, the
reflection with respect to this coordinate seems like the most natural
one. Nevertheless, the reflection symmetry with respect to this operator
imposed on an initial data for $E^3$, Nil or Sol model is {\it not
  preserved} under the time evolution. This fact can be seen directly
from the dynamical equations for the unbarred variables. For example,
those for the Nil model are given by (See Eq.\reff{eq:n-1})
\begin{eqnarray}
  \label{eq:d-3}
  \ddot P &-& e^{-2\tau}P''-e^{2P}(\dot Q^2-e^{-2\tau}(Q'-1)^2)=0,
  \nonumber \\
  \ddot Q &-& e^{-2\tau}Q''+2(\dot P\dot Q-e^{-2\tau}P'(Q'-1))=0,
\end{eqnarray}
which are apparently, in contrast to the $T^3$ model, not invariant
under the transformation $\vbrow x\goes - \vbrow x$, due to the factor
$(Q'-1)$. Similarly, those for the $E^3$ and Sol models are not
invariant under the same transformation. As a result, in contrast to
translation, there are no invariant subsets for reflection in the $E^3$,
Nil and Sol models which naturally correspond to the invariant subset
$\S(R_{0*})$ for the $T^3$ model.

However, the $T^3$ model admits another one-parameter family of
nontrivial reflection operators $R_{l_\theta *}$. (See
Eq.\reff{eq:sym-8-1-2}) The momentum part is derived from
Eq.\reff{eq:sym-8-1-2} by differentiating with respect to $\tau$ and using
Eqs.\reff{eq:h-1.5}. The result is given by
\begin{eqnarray}
  \label{eq:d-4}
  R_{l_\theta *} : \; (\bar\pi_P(x),\bar\pi_Q(x)) & \goes &
  \bigg( \frac{ 
    ( (\c+\bQ \s)^2 - e^{-2\bP}\s^2 ){\bar\pi}_P
    +2 e^{-2\bP} \s(\c+\bQ \s){\bar\pi}_Q }
  { (\c+\bQ \s)^2 +e^{-2\bP}\s^2 },
  \nonumber \\
  &&
  2\s(\c+\bQ\s){\bar\pi}_P
  - ((\c+\bQ\s)^2-e^{-2\bP}\s^2){\bar\pi}_Q
  \bigg)\bigg|_\shortPQ,
\end{eqnarray}
where $\c\equiv\cos 2\theta$, $\s\equiv\sin 2\theta$.

As shown in the previous subsection, for Nil and Sol models only
isolated values ($\theta=0$ or $\pi/2$ for Nil, and $\pi/4$ or $3\pi/4$
for Sol) of $\theta$ are permissible. As a result, the invariant subsets 
for reflection in the Nil and Sol models are much limited compared to
the $T^3$ model. More precisely, we have the following.
\begin{prop}
  Let $\S(R_T)$, $\S(R_E)$, $\S(R_N)$, and $\S(R_S)$ be the unions of
  the all invariant subsets for reflection in, respectively, the $T^3$,
  $E^3$, Nil, and Sol models. Then, the inclusion relation
\begin{equation}
  \label{eq:d-10}
  \S(R_T) \supset \S(R_E) \supset (\S(R_N),\; \S(R_S))
\end{equation}
holds (if neglecting the constraints corresponding to Eq.\reff{eq:a-5}).
\end{prop}

{\it Proof}: This can be seen from $\S(R_T)=
\S(R_{0*})\cup(\cup_\theta\S(R_{l_{\theta*}}))$, $\S(R_E)=
\cup_\theta\S(R_{l_{\theta*}})$, $\S(R_N)= \S(R_{l_{0*}})$, and
$\S(R_E)= \S(R_{l_{\pi/4*}})$. \endofproofmark

\medskip

It should be stressed that the relation \reff{eq:d-10} truly reflects
the dynamical properties of the models. Note that, since the (periodic)
boundary conditions imposed on initial data are the same for all models
if the phase space $\P$ is spanned by the unbarred variables, a map $f$
which preserves the boundary conditions for a model always preserves the
boundary conditions for other models, too. This means that there are the
same amounts of ``reflection symmetric initial data'' for the four
models. However, the time evolution does not always preserve the
symmetry of such an initial data, as we have seen. To conclude, we have
clarified how much ``reflection'' symmetric data exist {\it for which
  the symmetry is preserved} under the time evolution for each model,
and found, in particular, the inclusion relation \reff{eq:d-10}. This
may be interpreted as a manifestation of the influence of topology to
dynamics.

\section{Summary and Comments}
\label{sec:summary}

We have made the structures of \LUU-manifolds clear and classified the
possible topologies of them. For convenience we have split the set of
these manifolds into two kinds; {\it the first kind}, those with local
Killing vectors which are not degenerate everywhere, and {\it the second
  kind}, those with ones which are degenerate somewhere. The local
Killing vectors of any \LUU-manifold of the second kind are actually
defined globally, so that all the \LUUs\ models of the second kind are
contained in the usual Gowdy models. On the other hand, it is only $T^3$
that is contained in the Gowdy models among the varieties of the
\LUU-manifolds of the first kind, so that in this paper we have
basically restricted ourselves to the models of the first kind.

In the case that three-manifold $M$ is an \LUUs\ space of the first kind,
$M$ is a $T^2$-bundle over the $S^1$ (Theorem \ref{th:bundle}), and
according to the corresponding geometric structure, $M$ is naturally
characterized by one of $E^3$, Nil, or Sol. In each case, the possible
topologies of $M$ can be classified more precisely, but if we are
interested in the dynamical properties of the corresponding spacetime
models it is not necessary to consider all the models one by
one. Restricted to representatives for this ``dynamical equivalence
class,'' the number of models to consider is to great extent
decreased. In particular, there is three representatives for $E^3$
(cf. Sec.\ref{sec:e3}), and there is only one for Nil
(cf. Sec.\ref{sec:nil}). The representatives for Sol are parameterized
by only one discrete parameter $\q$ (cf. Sec.\ref{sec:sol}).

We have given two ways of representing the metric for each of $E^3$, Nil
and Sol models. (We distinguish between the $E^3$ and $T^3$ models. See
{\bf Convention} in Sec.\ref{sec:e3}.) Note that the metrics of all
the \LUUs\ spacetimes can be represented locally in the same canonical
form, but with distinct boundary conditions. In this view, however,
global symmetries that arise from the spatial topology tend to be
unclear. On the other hand, the other way of representing the metric
makes the geometric structure the \LUU-manifold $M$ admits
manifest. Geometrically it is obtained by ``relaxing'' the corresponding
locally homogeneous metric. This type of metric has another advantage
that it gives rise to a natural reduction of the model; that is, since
the boundary conditions are always given by the periodicity for the
spatial coordinate, the spatial manifold naturally reduces to the $S^1$,
the base space of the bundle. Note that this fact makes the
identifications of the phase spaces for the varieties of the models
possible.

Finally, as an application of these fundamental facts we have given the
translation and reflection operators which commute with the
time-evolution, and have discussed their significance. Since, as stated
above, the metrics of \LUUs\ models can always be locally represented in
the same form, these models are considered to have the same {\it local}
dynamical properties. However differences of topologies affect the {\it
  global} dynamical properties of the models through the differences of
the boundary conditions. In this paper we have examined whether or not a
global symmetry imposed on an initial data is preserved in time, or how
much there exist global symmetries which are preserved in time, for each
model of the first kind. As a result, we have indeed found that there
are remarkable distinctions in the properties of {\it reflection}. The
freedom of reflection symmetries is the largest as for the $T^3$ model,
and we have the inclusion relation \reff{eq:d-10}. In particular, naive
(even type) reflection symmetric initial data for the $E^3$, Nil or Sol
model do not evolve maintaining its symmetry in time, in contrast to the
$T^3$ model.

In the following we make some comments. A first one concerns another
point concerning the correlations between topology and dynamics. As for
the dynamics of the Gowdy $T^3$ model, what has been being taken an
interest most is whether the AVTD conjecture \cite{IM} is correct or not
that predicts a universal behavior of the approach toward the initial
singularity. This conjecture has been basically supported both
analytically \cite{GM,KR} and numerically \cite{BG,HS}. However, the
subtlety has been pointed out \cite{BG} that there is a measure-zero set
of spatial points where the AVTD behavior is not achieved. Here we
concern ourselves with these {\it nongeneric} points. It is known
\cite{BG} that at these points $\bQ'=0$. Since this condition is local,
it does not depend on the topology. However, note that while points of
$\bQ'=0$ are inevitable in the $T^3$ model, since $\bQ(x)$ is a (smooth)
periodic function, such points are not necessary in the Nil and Sol
models. Hence we can naturally expect that topology affects the tendency
of the appearance of the nongeneric points. The points where $\bQ'=0$
correspond, from Eqs.\reff{eq:n-1} and \reff{eq:s-10}, to the points
where
\begin{eqnarray}
  \label{eq:con-1-1}
  \mbox{Nil:} && \quad Q'= 1, \\
  \label{eq:con-1-2}
  \mbox{Sol:} && \quad (\log |Q|)' = 2\q,
\end{eqnarray}
for the unbarred variables. The unbarred $Q(x)$ must be periodic, but
this condition does not force the existence of points such that the
above conditions are fulfilled.  However, since the evolution of $Q$
freezes when approaching the initial singularity, we can naturally
expect that nongeneric points are still generated if the maximal
gradient of $Q$ at an initial surface is large enough and if, as a
result, $Q'_{\rm max}>1$ (Nil), or $(\log |Q|)'_{\rm max}>2\q$ (Sol) at
a time when the evolution of $Q$ freezes. Indeed, this property has been
observed from numerical simulations the author performed. A detailed
description for this point will be reported elsewhere. In Ref.\cite{T}
it was suggested that the nongeneric points are not generated in the Nil
and Sol models, since the conditions corresponding to
Eqs.\reff{eq:con-1-2} and \reff{eq:con-1-2} were written in such
different forms with different choice of metric functions that the
conditions seem not to be fulfilled.  This claim is not correct, as seen
from above. Still, we may claim that the $E^3$ model is most likely to
generate nongeneric points.

Next we comment on the locally $\U1$-symmetric models \cite{Mon} that
are less symmetric than the \LUUs\ models. This model admits only one
spatial local Killing vector. The spatial manifold $M$ is again assumed
to be closed. First, from an analogous analysis to the proof of Lemma
\ref{le:T2}, the Killing orbits are found to be closed and are
homeomorphic to $S^1$, if the local Killing vector does not degenerate
everywhere (that is, if the model is the ``first kind''). Hence $M$ is
naturally a Seifert fiber space \cite{Hem}, and still admits a geometric
structure, which is one of $E^3$, Nil, $H^2\times\R$, $\SL{2,\R}$,
$S^2\times\R$, or $S^3$ \cite{Sco}. In the case that $M$ is the ``second
kind'', i.e., the local Killing vector vanishes on somewhere, the
manifold $M$ is obtained from a Seifert fiber space by performing a
finite number of particular kind of Dehn surgeries. The resulting
manifold $M$ is found (from one in the series of famous theorems of
Thurston concerning the $H^3$ structure) to again admit a geometric
structure. Detailed accounts and applications will be presented
elsewhere.

\appendix
\section{Vacuum Einstein equations for Gowdy Spacetimes}
\label{sec:gc}

In this Appendix, we summarize the standard prescription \cite{Gow2} of
the reduced vacuum Einstein equations for Gowdy spacetimes, together
with generalizations to Nil and Sol types.

We consider the two-surface orthogonal metric \reff{eq:g1metric}. This
is used most frequently in the literature since the Einstein equations
become to great extent simpler.

More explicitly, we write \cite{BM}
\begin{equation}
  \label{eq:c-metric}
  \d s^2= e^{\bar\gamma/2}(-\d t^2+\d x^2)+
  \bR (e^{\bP}(\d y+\bQ\d z)^2+e^{-\bP}\d z^2).
\end{equation}
where $\bP$ and $\bQ$ are functions of $t$ and $x$. Then, the vacuum
Einstein equations for the metric functions are explicitly given as
follows. The dynamical equations are
\begin{eqnarray}
  \label{eq:dyn}
  \bP_{tt} &-& \bP''-e^{2\bP}(\bQ_t^2-\bQ'{}^2)
  +\bR\inv (\bR_t \bP_t-\bR'\bP')=0, \\ 
  \bQ_{tt} &-& \bQ''+2(\bP_t \bQ_t-\bP'\bQ')
  +\bR\inv (\bR_t \bQ_t-\bR'\bQ')=0, \nonumber
\end{eqnarray}
and
\begin{equation}
    \label{eq:dynR}
  \bR_{tt} - \bR''=0,
\end{equation}
where subscript $t$ represents the derivative with respect to $t$ and
the primes represent the derivatives with respect to $x$. The remaining
constraint equations takes a simple form in the null coordinates $u=t-x$
and $v=t+x$;
\begin{eqnarray}
  \label{eq:ctr}
  \bR(\bP_u^2+e^{2\bP}\bQ_u^2)\wa -2\bR_{uu}+ \bR_u (\bgamma_u +\bR\inv \bR_u),
  \nonumber \\
  \bR(\bP_v^2+e^{2\bP}\bQ_v^2)\wa -2\bR_{vv}+ \bR_v (\bgamma_v +\bR\inv \bR_v),
\end{eqnarray}
where the subscripts stand for the derivatives thereof. 

The last two equations can be solved for $\bgamma_u$ and $\bgamma_v$ at
any spacetime point wherever $\bR_u$ and $\bR_v$ do not vanish. In this
case the integrability condition for $\bgamma$, namely
$\bgamma_{uv}-\bgamma_{vu}=0$, is automatically satisfied if the
dynamical equations \reff{eq:dyn} are satisfied. Hence we can integrate
the equations and obtain $\bgamma$ by using solutions of the
(unconstrained) dynamical equations \reff{eq:dyn}. If there is a
spacetime point such that $\bR_u=0$ (or $\bR_v=0$), Eqs.\reff{eq:ctr}
cannot be solved for $\bgamma_u$ (or $\bgamma_v$), so that the constraint
equation \reff{eq:ctr} constrains $\bP$ and $\bQ$ at that spacetime
point. This condition is called the ``matching condition''
\cite{Gow2}. Explicitly,
\begin{eqnarray}
  \label{eq:mc}
  \bP_u^2+e^{2\bP}\bQ_u^2\wa -2\bR\inv \bR_{uu}\quad {\rm at}\quad \bR_u=0,
  \nonumber \\
  \bP_v^2+e^{2\bP}\bQ_v^2\wa -2\bR\inv \bR_{vv}\quad {\rm at}\quad \bR_v=0.
\end{eqnarray}

From the observation that the left hand sides of Eqs.\reff{eq:mc} are
positive semidefinite, we find that for $R>0$,
\begin{eqnarray}
  \label{eq:semi}
  \bR_{uu}\leq 0 \quad {\rm at}\quad \bR_u=0, \nonumber \\
  \bR_{vv}\leq 0 \quad {\rm at}\quad \bR_v=0.
\end{eqnarray}
An implication of these inequalities is called the ``corner theorem''
\cite{Gow2}. Here, a {\it corner} is a point in the $t$-$x$ plane such
that $\bR_u=0$ or $\bR_v=0$, that is, a point where the gradient of a
level curve of $\bR$ becomes null. The implication of the inequalities
would be clear if observing some level curves of $\bR$ near a corner in
the $u$-$v$ coordinates. It should be stressed that these inequalities
do {\it not} depend on the boundary conditions for $\bP$, $\bQ$, and
$\bgamma$, so hold for any spatial topology.  (This theorem, however,
tacitly assume the finiteness of $\bgamma_u$ and $\bgamma_v$ at the
corner. If we allow for $\bR_u\bgamma_u$ and $\bR_v\bgamma_v$ to remain
finite at the corner with diverging $\bgamma_u$ and $\bgamma_v$, we may
have counterexamples to this theorem. \cite{Tomita})

For any \LUU-symmetric spacetime of the first kind, the area function
$\bR$ must be periodic (cf. Eq.\reff{eq:g1bc}). Hence, corners appear
only in pair, implying that $\bR$ cannot have a corner. We can therefore
choose $\bR$ spatially constant
\begin{equation}
  \label{eq:a-1}
  \bR(t,x)=t,
\end{equation}
(using the remaining freedom of the coordinate transformations of type
$u\goes F(u),v\goes G(v)$).  In this case, the reduced Einstein
equations \reff{eq:dyn} and \reff{eq:ctr} become
\begin{eqnarray}
  \label{eq:a-2}
  \ddot \bP &-& e^{-2\tau}\bP''-e^{2\bP}(\dot \bQ^2-e^{-2\tau}\bQ'{}^2)=0, \\
  \ddot \bQ &-& e^{-2\tau}\bQ''+2(\dot \bP\dot \bQ-e^{-2\tau}\bP'\bQ')=0,
  \nonumber
\end{eqnarray}
with
\begin{eqnarray}
  \label{eq:a-3}
  \dot{\bar\lambda} \wa
  \dot \bP^2+e^{-2\tau}\bP''+e^{2\bP}(\dot \bQ^2+e^{-2\tau}\bQ'{}^2), \\
  \bar\lambda' \wa 2( \bP'\dot \bP+e^{2\bP}\bQ'\dot \bQ ).
  \nonumber
\end{eqnarray}
Here, we have put
\begin{equation}
  \label{eq:a-4}
  t=e^{-\tau},\quad e^{\bgamma/2}=e^{-\blambda/2+\tau/2}.
\end{equation}
Dots in these equations represent derivatives with respect to $\tau$
(not $t$). $\bar\lambda$ should be periodic for any \LUU-symmetric
spacetime of the first kind. This leads to the constraint for $\bP$ and
$\bQ$,
\begin{equation}
  \label{eq:a-5}
  \int_0^{2\pi}(\bP'\dot \bP+e^{2\bP}\bQ'\dot \bQ)\d x=0,
\end{equation}
because of the second equation of Eqs.\reff{eq:a-3}.

The functions $\bP$ and $\bQ$ are not periodic functions for the Nil and 
Sol models. Appropriate boundary conditions are obtained from
Eqs.\reff{eq:n-1} and \reff{eq:s-10}. Specifically,
\par\noindent
For Nil:
\begin{equation}
  \label{eq:a-6}
  \bP(x+2\pi)=\bP(x),\; \bQ(x+2\pi)=\bQ(x)-2\pi.
\end{equation}
For Sol:
\begin{equation}
  \label{eq:a-7}
  \bP(x+2\pi)=\bP(x)+4\pi\q,\; \bQ(x+2\pi)=e^{-4\pi\q}\bQ(x).
\end{equation}

\remarkmark The last boundary conditions \reff{eq:a-6} and \reff{eq:a-7}
do not necessarily coincide with those obtained from Eqs.\reff{eq:g1bc}.
This is because the choice of the covering group $\Gamma$
(cf. Eq.\reff{eq:covmap}) is different. In obtaining Eqs.\reff{eq:g1bc}
we have fixed $\Gamma$ first and found the appropriate metric, but for
Eqs.\reff{eq:a-6} and \reff{eq:a-7} we try to fix the universal covering
metric with appropriate $\Gamma$ as long as possible.  This is the
essential point in the dynamical equivalences for the Nil and Sol
models presented in Sec.\ref{sec:reduction}.

\section{Compactification of Sol}
\label{sec:comsol}

In this Appendix we show calculations needed to perform the
compactification of Sol. This is basically a review of that presented in
Ref.\cite{KTH}. In this reference, the form of $\omcg{T^2}\simeq
\SL{2,\Z}$ was restricted to that of Eq.\reff{eq:Asol}, and the
parameter $\q$ (See bellow) was not introduced. In this review we show
an explicit calculation with general form of the matrix and with $\q$,
though no essential difference appears.

What we do is to embed the fundamental groups into Sol$\simeq\Gsix$ (the
Bianchi ${\rm VI}_0$ group).  For notational convenience, letting
$a=g_2$, $b=g_3$, and $c=g_1$ in the representation \reff{eq:pi1},
\begin{equation}
  \label{eq:pi1sol}
  \pi_1=\angl{a,b,c;\; [a,b]=1,cac\inv=a^p b^r,
    cbc\inv=a^q b^s},
\end{equation}
where
\begin{equation}
  \label{eq:apsol-1}
  A\equiv\smatrix pqrs\in\SL{2,\Z},\quad \mbox{with}\quad \abs{\Tr A}>2.
\end{equation}
We represent the $\pi_1$-generators with their components in $\Gsix$
\begin{equation}
  \label{eq:apsol-2}
  a=\vector{a_1}{a_2}{a_3},\;b=\vector{b_1}{b_2}{b_3},\;
  c=\vector{c_1}{c_2}{c_3}.
\end{equation}
To avoid confusions with powers, we use subscripts to distinguish each
components as above. The multiplication rule for $\Gsix$ is given by
\begin{equation}
\label{eq:apsol-3}
\vector\alpha\beta\gamma \vector xyz=
\vector{\alpha+x}{\beta+e^{-\q\alpha}y}{\gamma+e^{\q\alpha}z},\quad
\vector\alpha\beta\gamma\inv=
\vector{-\alpha}{-e^{\q\alpha}\beta}{-e^{-\q\alpha}\gamma},
\end{equation}
where $\q>0$ is a parameter (, which has nothing to do with the elements
in the matrix $A$). Left invariant one-forms are then given by
Eq.\reff{eq:1-sol}.

First, note the first components of the relations
\begin{equation}
  \label{eq:apsol-4}
  cac\inv=a^pb^r,\; cbc\inv=a^qb^s.
\end{equation}
Noting that the first component of the product of two elements of Sol
is simply given by the sum of the first components of the elements, we
obtain $a_1=pa_1+rb_1$ and $b_1=qa_1+sb_1$, that is,
\begin{equation}
  \label{eq:apsol-5}
  \smatrix{p-1}rq{s-1}\svector{a_1}{b_1}=\svector 00.
\end{equation}
The determinant for the matrix appearing in the left hand side is given
by $\det A-\Tr A=1-\Tr A$, which is less than $-1$ (when $\Tr A>2$) or
greater than $3$ (when $\Tr A<-2$), from the assumptions on the matrix
$A$. Hence, the matrix has its inverse, implying
\begin{equation}
  \label{eq:apsol-6}
  a_1=b_1=0.
\end{equation}
Under this condition the relation $[a,b]=1$ becomes trivial. Moreover,
we can at once calculate
\begin{equation}
  \label{eq:apsol-7}
  a^p=\vector0{pa_2}{pa_3},\; b^r=\vector0{rb_2}{rb_3},\;
  cac\inv=\vector0{a_2e^{-\q c_1}}{a_3e^{ \q c_1}},\; \mbox{etc.},
\end{equation}
so that the second and third components of the relations
\reff{eq:apsol-4} are given by
\begin{equation}
  \label{eq:apsol-8}
  A\svector{a_2}{b_2}=e^{\q c_1}\svector{a_2}{b_2},\quad
  A\svector{a_3}{b_3}=e^{-\q c_1}\svector{a_3}{b_3}.
\end{equation}
These reveal that $\svector{a_2}{b_2}$ and $\svector{a_3}{b_3}$ are
eigenvectors of $A$, and $e^{\pm\q c_1}$ are the eigenvalues. The
characteristic polynomial is $\lambda^2-\Tr A\,\lambda+\det A=0$, or
equivalently,
\begin{equation}
  \label{eq:apsol-9}
  \lambda^2-\Tr A\,\lambda+1=0.
\end{equation}

Since the eigenvalues $e^{\pm\q c_1}$ must be positive, the embedding is
found to be possible only for the case $\Tr A>2$. However, even for the
case $\Tr A<-2$, if the map
\begin{equation}
  \label{eq:apsol-10}
  h:\; (x,y,z)\goes (x,-y,-z)
\end{equation}
can be regarded as an isometry, we can perform the embedding by putting
$ c=h\circ\vector{c_1}{c_2}{c_3}$, where the column vector in the right
hand side is an element of Sol, and $\circ$ represents the composition
of maps. It is an easy task to retrace the calculation above. We will
find in particular that the eigenvalues of $A$ must equal to $-e^{\pm \q
  c_1}$ rather than $e^{\pm \q c_1}$.

Finally, the formal solutions for embedding are as follows.

(i) Case of $\Tr A>2$: 
Letting the eigensystems (the pairs of an eigenvalue and the
corresponding normalized eigenvector) for the matrix $A$ be
\begin{equation}
  \label{eq:apsol-12}
  \brace{e^{\q c_1},\svector{u_2}{u_3}},\quad 
  \brace{e^{-\q c_1},\svector{v_2}{v_3}},
\end{equation}
the solution is
\begin{equation}
  \label{eq:apsol-13}
  a=\vector0{\beta u_2}{\gamma v_2},\;
  b=\vector0{\beta u_3}{\gamma v_3},\;
  c=\vector{c_1}{c_2}{c_3},
\end{equation}
where $\beta$, $\gamma$, $c_2$, and $c_3$ are parameters. Note that,
while the product $q c_1$ is directly connected with the eigenvalues of
$A$, we can freely specify one of $q$ or $c_1$.

(ii) Case of $\Tr A<-2$:
Letting the eigensystems be
\begin{equation}
  \label{eq:apsol-14}
  \brace{-e^{\q c_1},\svector{u_2}{u_3}},\quad 
  \brace{-e^{-\q c_1},\svector{v_2}{v_3}},
\end{equation}
the solution is
\begin{equation}
  \label{eq:apsol-15}
  a=\vector{\beta u_2}{\gamma v_2}0,\;
  b=\vector{\beta u_3}{\gamma v_3}0,\;
  c=h\circ\vector{c_1}{c_2}{c_3},
\end{equation}
where $\beta$, $\gamma$, $c_2$ and $c_3$ are parameters. As in the case
(i), one of $q$ or $c_1$ can be freely specified.

\medskip

\remarkmark We have seen that the fundamental group for $\Tr A>2$ can be
embedded into $\Gsix$. Also, we have seen that the fundamental group for
$\Tr A< -2$ can be embedded into $\GsixZ$, the group generated by
$\Gsix$ and the $\Z_2$-factor $h$. It is easy to see that the same
groups can be embedded into the smaller group $\Hsix\subset \Gsix$ which
is given by making the first component of $\Gsix$ discrete like $c_1\Z$,
or into $\HsixZ$, where $\HsixZ\subset\GsixZ$ is the group generated by
$\Hsix$ and the map $h$. Since $\HsixZ$ is the isometry group of the
universal covering spacetime \reff{eq:sol-metric} (with the appropriate
choice of $\q$, as shown in Sec.\ref{sec:sol}), the same embeddings are
possible for this (inhomogeneous) spacetime.

\section*{Acknowledgments}
The author thanks Professor K.~Tomita for explaining a subtle point in
the Gowdy equations to him.  He acknowledges financial support from the
Japan Society for the Promotion of Science.


\end{document}